\documentclass[12pt]{article}
\pdfoutput=1
\usepackage[a4paper,left=25mm,top=25mm,textwidth=160mm,textheight=249.7mm]{geometry}
\usepackage{amsmath,amsthm,amssymb}
\usepackage[utf8]{inputenc}
\usepackage[english]{babel}
\usepackage{hyperref} 
\usepackage{leftidx}
\usepackage{enumerate}
\usepackage[pdftex]{graphicx}
	\DeclareGraphicsExtensions{.pdf,.png,.jpg,.bmp}
 \usepackage{float}
 \usepackage{subcaption}
\usepackage{longtable}

\usepackage{graphicx}
\usepackage[labelformat=simple]{subcaption}

\DeclareCaptionLabelFormat{subcaptionlabel}{\normalfont(\textbf{#2}\normalfont)}
\usepackage{longtable}
\usepackage{booktabs}

\begin{document}

\title{An Accelerating Universe without Lambda: Delta~Gravity Using Monte Carlo }
\author{ Jorge Alfaro\footnote{Instituto de F\'isica, Pontificia Universidad Cat\'olica de Chile. Casilla 306, Santiago, Chile. jalfaro@uc.cl.}, Marco San Martín\footnote{Instituto de Astronom\'ia, Pontificia Universidad Cat\'olica de Chile. Casilla 306, Santiago, Chile. mlsanmartin@uc.cl },Joaquín Sureda\footnote{Instituto de Astronom\'ia, Pontificia Universidad Cat\'olica de Chile. Casilla 306, Santiago, Chile. jmsureda@uc.cl}}
 
\maketitle

\begin{abstract}

A gravitational field model based on two symmetric tensors, $g_{\mu \nu}$ and $\tilde{g}_{\mu \nu}$, is studied, using a Markov Chain Monte Carlo (MCMC) analysis with the most updated catalog of SN-Ia. In this model, new matter fields are added to the original matter fields, motivated by an additional symmetry ($\tilde{\delta}$~symmetry). We call them $\tilde{\delta}$ matter fields. This theory predicts an accelerating Universe without the need to introduce a cosmological constant $\Lambda$ by hand in the equations.
We obtained a very good fit to the SN-Ia Data, and with this, we found the two free parameters of the theory called $C$ and $L_2$. With these values, we have fixed all the degrees of freedom in the model.
The last $H_0$  local value measurement is in tension with the CMB Data from Planck.  Based on an absolute magnitude $M_V = -19.23$ for the SN, Delta Gravity finds $H_0$ to be $74.47\pm 1.63$ km/(s Mpc). This value is in concordance with the last measurement of the $H_0$ local value, $73.83\pm 1.48$ km/(s Mpc).
\end{abstract}

\section{Introduction}

General relativity (GR) is valid on scales larger than a millimeter to the solar-system scale
~\cite{GR_scale_1, GR_scale_2}. Nevertheless,~the~theory is non-renormalizable, which prevents its unification with the other forces of nature. Trying~to quantize GR is the main physical motivation of string theories~\cite{string1,string2}. Moreover, recent~discoveries in cosmology~\cite{Weinberg,Riess1998,DM_DE3,DM_DE4} have revealed that most part of matter is in the form of unknown matter, dark matter (DM), and~that the dynamics of the expansion of the Universe is governed by a mysterious component that accelerates the expansion, Dark Energy (DE). Although~GR can accommodate both DM and DE, the~interpretation of the dark sector in terms of fundamental theories of elementary particles is problematic.

Although some candidates exist that could play the role of DM, none have been detected yet. Also,~an~alternative explanation based on the modification of the dynamics for small accelerations cannot be ruled out~\cite{DM_DE6, DM_DE7}. On~the other side, DE can be explained if a small cosmological constant ($\Lambda$) is present. In~early times after the Big Bang, this constant is irrelevant, but~at the later stages of the evolution of the Universe $\Lambda$ will dominate the expansion, explaining the acceleration. Such~small $\Lambda$ is very difficult to generate in Quantum Field Theory (QFT) models, because $\Lambda$ is the vacuum energy, which is usually very large~\cite{DM_DE5}.

One of the most important mysteries in cosmology and cosmic structure formation is to understand the nature of Dark Energy in the context of a fundamental physical theory~\cite{DE1, DE2}. In~recent years there has been various proposals to explain the observed acceleration of the Universe. They~involve the inclusion of some additional fields in approaches such as quintessence, chameleon, vector DE, or massive gravity; The addition of higher order terms in the Einstein-Hilbert action, such as $f(R)$ theories and Gauss-Bonnet terms and finally the introduction of extra dimensions for a modification of gravity on large scales (See~\cite{DE3}).

Other interesting possibilities are~the search for non-trivial ultraviolet fixed points in gravity (asymptotic safety~\cite{GR_Weinberg}) and the notion of induced gravity~\cite{induced_gravity1,induced_gravity2,induced_gravity3,induced_gravity4}. The~first possibility uses exact renormalization-group techniques~\cite{Ren_group_1,Ren_group_4} together with lattice and numerical techniques such as Lorentzian triangulation analysis~\cite{Lorentz_triang}. Induced~gravity proposes that gravitation is a residual force produced by other interactions.

Recently, in~\cite{Alfaro_0, Alfaro_1} a field theory model explores the emergence of geometry by the spontaneous symmetry breaking of a larger symmetry where the metric is absent. Previous~work in this direction can be found in~\cite{work_1,work_7}.

In a previous work~\cite{DeltaGravityL}, we~studied a model of gravitation that is very similar to classical GR, but~could make sense at the quantum level. In~the construction, we~consider two different points. The~first is that GR is finite on shell at one loop~\cite{tHooft}, so~renormalization is not necessary at this level. The~second is a type of gauge theories, $\tilde{\delta}$ Gauge Theories (Delta Gauge Theories), presented in~\cite{DGT1,DGT2}, which main properties are: (a) New kind of fields are created, $\tilde{\phi}_I$, from the originals $\phi_I$. (b) The classical equations of motion of $\phi_I$ are satisfied in the full quantum theory. (c) The model lives at one loop. (d)~The action is obtained through the extension of the original gauge symmetry of the model, introducing an extra symmetry that we call $\tilde{\delta}$ symmetry, since it is formally obtained as the variation of the original symmetry. When~we apply this prescription to GR we obtain Delta Gravity. Quantization~of Delta Gravity is discussed in~\cite{Delta_Gravity}.

Here, we~study the classical effects of Delta Gravity at the cosmological level. For~this, we~assume that the Universe is composed by non-relativistic matter (DM, baryonic matter) and radiation (photons, massless particles), which satisfy a fluid-like equation $p = \omega \rho$. Matter~dynamics is not considered, except by demanding that the energy-momentum tensor of the matter fluid is covariantly conserved. This~is required to respect the symmetries of the model. In~contrast to~\cite{DG_DE}, where an approximation is discussed, in~this work we find the exact solution of the equations corresponding to the above suppositions. This~solution is used to fit the SN-Ia Data and we obtain an accelerated expansion of the Universe in the model without DE.

It was noticed in~\cite{DGT2} that the Hamiltonian of delta models is not bounded from below. Phantoms~cosmological models~\cite{Robert2,Robert1} also have this property. Although~it is not clear whether this problem will subsist or not in a diffeomorphism-invariant model as Delta Gravity. Phantom~fields are used to explain the expansion of the Universe. Then, even if it could be said that our model works on similar grounds, the~accelerated expansion of the Universe is really produced by a constant $L_2 \neq 0$ (it~is a integration constant that comes from the Delta Field Equations), not~by a phantom field.

It should be remarked that Delta Gravity is not a metric model of gravity because massive particles do not move on geodesics. Only~massless particles move on null geodesics of a linear combination of both tensor fields.

\section{Definition of Delta Gravity}

In this section, we define the action as well as the symmetries of the model
and derive the equations of motion.

%----nuevo!!
These modified theories consist in the application of a variation represented by $\tilde{\delta}$. As~a variation, it~will have all the properties of a usual variation such as:
\begin{eqnarray}
\tilde{\delta}(AB)&=&\tilde{\delta}(A)B+A\tilde{\delta}(B) \nonumber \\
\tilde{\delta}\delta A &=&\delta\tilde{\delta}A \\
\tilde{\delta}(\Phi_{, \mu})&=&(\tilde{\delta}\Phi)_{, \mu} \nonumber
\end{eqnarray}
where $\delta$ is another variation. The~particular point with this variation is that, when we apply it on a field (function, tensor, etc.), it~will give new elements that we define as $\tilde{\delta}$ fields, which is an entirely new independent object from the original, $\tilde{\Phi} = \tilde{\delta}(\Phi)$. We~use the convention that a tilde tensor is equal to the $\tilde{\delta}$ transformation of the original tensor when all its indexes are covariant.

First, we~need to apply the $\tilde{\delta}$ prescription to a general action. The~extension of the new symmetry is given by:

\begin{equation}
S_0 =\int d^n x \mathcal{L}_0(\phi,\partial_i \phi) \to S = \int d^n x \left(\mathcal{L}_0(\phi,\partial_i \phi)+ \tilde{\delta} \mathcal{L}_0(\phi,\partial_i \phi)\right)
\end{equation}
where $S_0$ is the original action, and~$S$ is the extended action in Delta Gauge Theories.

GR is based on Einstein-Hilbert action, then,

\begin{equation}
S_0 =\int d^4 x \mathcal{L}_0(\phi)= \int d^4 x \sqrt{-g} \left(\frac{R}{2\kappa}+L_M \right)
\end{equation}
where $L_M = L_M(\phi_I,\partial_{\mu}\phi_I)$ is the Lagrangian of the matter fields $\phi_I$, $\kappa = \frac{8 \pi G}{c^2}$. Then, the Delta Gravity action is given by,

\begin{eqnarray}
\label{grav action}
S =S_0 + \tilde{\delta} S_0= \int d^4x \sqrt{-g} \left(\frac{R}{2\kappa} + L_M - \frac{1}{2\kappa}\left(G^{\alpha \beta} - \kappa T^{\alpha \beta}\right)\tilde{g}_{\alpha \beta} + \tilde{L}_M\right)
\end{eqnarray}
where we have used the definition of the new symmetry: $\tilde{\phi} = \tilde{\delta}\phi$ and the metric convention of{~\cite{Weinberg}}\footnote{In~\cite{ag} can be found more about the formalism of the Delta Gravity action and the new symmetry $\tilde{\delta}$.}.

%---------------------------
Here:
\begin{eqnarray}
\tilde{g}_{\mu \nu} = \tilde{\delta} g_{\mu \nu},
\label{EM Tensor}
T^{\mu \nu} = \frac{2}{\sqrt{-g}} \frac{\delta\left(\sqrt{-g} L_M\right)}{\delta g_{\mu \nu}} \\
\label{tilde L matter}
\tilde{L}_M = \tilde{\phi}_I\left(\frac{\delta L_M}{\delta \phi_I}\right) + (\partial_{\mu}\tilde{\phi}_I)\left(\frac{\delta L_M}{\delta (\partial_{\mu}\phi_I)}\right)
\end{eqnarray}
and $\tilde{\phi}_I = \tilde{\delta}\phi_I$ are the $\tilde{\delta}$ matter fields. Then, the equations of motion are:
\begin{eqnarray}
\label{Einst Eq} G^{\mu \nu} &=& \kappa T^{\mu \nu} \\
\label{tilde Eq} F^{(\mu \nu) (\alpha \beta) \rho
\lambda} D_{\rho} D_{\lambda} \tilde{g}_{\alpha \beta} + \frac{1}{2}g^{\mu \nu}R^{\alpha \beta}\tilde{g}_{\alpha \beta} - \frac{1}{2}\tilde{g}^{\mu \nu}R &=& \kappa\tilde{T}^{\mu \nu}
\end{eqnarray}
with:
\begin{eqnarray}
\label{F}
F^{(\mu \nu) (\alpha \beta) \rho \lambda} &=& P^{((\rho
\mu) (\alpha \beta))}g^{\nu \lambda} + P^{((\rho \nu) (\alpha
\beta))}g^{\mu \lambda} - P^{((\mu \nu) (\alpha \beta))}g^{\rho
\lambda} - P^{((\rho \lambda) (\alpha \beta))}g^{\mu \nu} \nonumber \\
P^{((\alpha \beta)(\mu \nu))} &=& \frac{1}{4}\left(g^{\alpha
\mu}g^{\beta \nu} + g^{\alpha \nu}g^{\beta \mu} - g^{\alpha
\beta}g^{\mu \nu}\right) \nonumber\\
\tilde{T}^{\mu \nu}&=&\tilde{\delta} T^{\mu \nu} \nonumber
\end{eqnarray}
where $(\mu \nu)$ denotes that $\mu$ and $\nu$ are in a totally symmetric combination. An~important fact to notice is that our equations are of second order in derivatives which is needed to preserve causality. We~can show that $\eqref{tilde Eq}_{\mu \nu} = \tilde{\delta}\left[\eqref{Einst Eq}_{\mu \nu}\right]$. The~action \eqref{grav action} is invariant under \eqref{trans g} and \eqref{trans gt} (extended general coordinate transformations), given by:
\begin{eqnarray}
\label{trans g}
\bar{\delta} g_{\mu \nu} &=& \xi_{0 \mu ; \nu} + \xi_{0 \nu ; \mu} \\
\label{trans gt}
\bar{\delta} \tilde{g}_{\mu \nu} ( x ) &=& \xi_{1 \mu ; \nu} + \xi_{1\nu ; \mu} + \tilde{g}_{\mu \rho} \xi_{0, \nu}^{\rho} + \tilde{g}_{\nu \rho} \xi^{\rho}_{0, \mu} + \tilde{g}_{\mu \nu,\rho} \xi_0^{\rho}
\end{eqnarray}

This means that two conservation rules are satisfied. They~are:
\begin{eqnarray}
\label{Conserv T}
D_{\nu}T^{\mu \nu} &=& 0 \\
\label{Conserv tilde T}
D_{\nu}\tilde{T}^{\mu \nu} &=& \frac{1}{2}T^{\alpha \beta}D^{\mu}\tilde{g}_{\alpha \beta} - \frac{1}{2}T^{\mu \beta} D_{\beta}\tilde{g}^{\alpha}_{\alpha} + D_{\beta}(\tilde{g}^{\beta}_{\alpha}T^{\alpha \mu})
\end{eqnarray}

It is easy to see that \eqref{Conserv tilde T} is $\tilde{\delta}\left(D_{\nu}T^{\mu \nu}\right) = 0$.

%%%%%Test Particle.%%%%%
\section{Particle Motion in the Gravitational Field}

We are aware of the presence of the gravitational field through its effects on
test particles. For~this reason, here we discuss the coupling of a test
particle to a background gravitational field, such that the action of the
particle is invariant under \eqref{trans g} and \eqref{trans gt}.

In Delta Gravity we postulate the following action for a test particle:
\begin{eqnarray}
S_p = m \int \frac{dt}{\sqrt{- g_{\alpha \beta} \dot{x}^{\alpha}
\dot{x}^{\beta}}} \left( g_{\mu \nu} + \frac{1}{2} \tilde{g}_{\mu
\nu} \right) \dot{x}^{\mu} \dot{x}^{\nu} \label{geo}
\end{eqnarray}

Notice that $S_p$ is invariant under \eqref{trans g} and
$t$-parametrizations.

Since far from the sources, we~must have free particles in Minkowski space, i.e.,
$g_{\mu \nu} \sim \eta_{\mu \nu}, \tilde{g}_{\mu \nu}\sim 0$, it
follows that we are describing the motion of a particle of mass $m$.
Moreover, all massive particles fall with the same acceleration.

To include massless particles, we~prefer to use the action
{~\cite{Siegel}}:
\begin{eqnarray}
L = \frac{1}{2} \int dt \left( vm^2 - v^{- 1} \left( g_{\mu \nu} +
\tilde{g}_{\mu \nu} \right) \dot{x}^{\mu} \dot{x}^{\nu} +
\frac{m^2 + v^{- 2} \left( g_{\mu \nu} + \tilde{g}_{\mu \nu}
\right) \dot{x}^{\mu} \dot{x}^{\nu}}{2 v^{- 3} g_{\alpha \beta}
\dot{x}^{\alpha} \dot{x}^{\beta}} \left( m^2 + v^{- 2} g_{\lambda \rho}
\dot{x}^{\lambda} \dot{x}^{\rho} \right) \right) \label{geo3}
\end{eqnarray}

This action is invariant under reparametrizations:
\begin{equation}
x' (t') = x (t) ;\quad dt' v' (t') = dtv (t) ;\quad t' = t - \varepsilon (t)
\label{par}
\end{equation}

The equation of motion for $v$ is:
\begin{eqnarray}
v = - \frac{\sqrt{- g_{\mu \nu} \dot{x}^{\mu} \dot{x}^{\nu}}}{m} \label{v
eq}
\end{eqnarray}

Replacing \eqref{v eq} into \eqref{geo3}, we~get back \eqref{geo}.

Let us consider first the massive case. Using~\eqref{par} we can fix the gauge
$v = 1$. Introducing~$mdt = d \tau$, we~get the action:
\begin{eqnarray}
\mathcal{L}_1 = \frac{1}{2} m \int d \tau \left( 1 - \left( g_{\mu \nu} +
\tilde{g}_{\mu \nu} \right) \dot{x}^{\mu} \dot{x}^{\nu} + \frac{1 + \left(
g_{\mu \nu} + \tilde{g}_{\mu \nu} \right) \dot{x}^{\mu}
\dot{x}^{\nu}}{2 g_{\alpha \beta} \dot{x}^{\alpha} \dot{x}^{\beta}}
\left( 1 + g_{\lambda \rho} \dot{x}^{\lambda} \dot{x}^{\rho} \right)
\right) \label{geomassive}
\end{eqnarray}
plus the constraint obtained from the equation of motion for $v$:
\begin{equation}
g_{\mu \nu} \dot{x}^{\mu} \dot{x}^{\nu} = - 1 \label{shell}
\end{equation}

From $\mathcal{L}_1$ the equation of motion for massive particles is derived. We
define: $\overline{\mathfrak{g}}_{\mu \nu} = g_{\mu \nu} + \frac{1}{2}
\tilde{g}_{\mu \nu}$.
\begin{eqnarray}
\frac{d ( \dot{x}^{\mu} \dot{x}^{\nu} \overline{\mathfrak{g}}_{\mu \nu}
\dot{x}^{\beta} g_{\alpha \beta} + 2 \dot{x}^{\beta}
\mathfrak{\bar{g}}_{\alpha \beta})}{d \tau} - \frac{1}{2} \dot{x}^{\mu}
\dot{x}^{\nu} \mathfrak{\bar{g}}_{\mu \nu} \dot{x}^{\beta}
\dot{x}^{\gamma} g_{\beta \gamma, \alpha} - \dot{x}^{\mu} \dot{x}^{\nu}
\overline{\mathfrak{g}}_{\mu \nu, \alpha} = 0 & & \label{geo2}
\end{eqnarray}

The motion of massive particles is discussed in~\cite{NRAG}.

The action for massless particles is:
\begin{eqnarray}
\mathcal{L}_0 = \frac{1}{4} \int dt \left( - v^{- 1} \left( g_{\mu \nu} +
\tilde{g}_{\mu \nu} \right) \dot{x}^{\mu} \dot{x}^{\nu} \right)
\label{massless}
\end{eqnarray}

In the gauge $v = 1$, we~get:
\begin{eqnarray}
\mathcal{L}_0 = - \frac{1}{4} \int dt \left( g_{\mu \nu} + \tilde{g}_{\mu
\nu} \right) \dot{x}^{\mu} \dot{x}^{\nu} \label{massless2}
\end{eqnarray}
plus the equation of motion for $v$ evaluated at $v = 1$: $\left( g_{\mu \nu}
+ \tilde{g}_{\mu \nu} \right) \dot{x}^{\mu} \dot{x}^{\nu} = 0$. Therefore,~the~massless particle moves in a null geodesic of $\mathfrak{g}_{\mu \nu}= g_{\mu \nu} + \tilde{g}_{\mu \nu}$.

\section{Distances and Time Intervals}

In this section, we~define the measurement of time and distances in the model.

In GR the geodesic equation preserves the proper time of the particle along
the trajectory. \ Equation~\eqref{geo2} satisfies the same property: Along the
trajectory $\dot{x}^{\mu} \dot{x}^{\nu} g_{\mu \nu}$ is constant. Therefore~we
define proper time using the original metric $g_{\mu \nu}$,
\begin{equation}
d \tau = \sqrt{- g_{\mu \nu} dx^{\mu} dx^{\nu}} = \sqrt{- g_{00}} dx^0, \quad
(dx^i = 0) \label{proper}
\end{equation}

Following {~\cite{Landau}}, we~consider the motion of light rays along
infinitesimally near trajectories and~(\ref{proper}) to get the three
dimensional metric:
\begin{eqnarray}
dl^2 = \gamma_{ij} dx^i dx^j & & \nonumber\\
\gamma_{ij} = \frac{g_{00}}{\mathfrak{g}_{00}} \left( \mathfrak{g}_{ij} -
\frac{\mathfrak{g}_{0 i} \mathfrak{g}_{0 j}}{\mathfrak{g}_{00}}\right) & &
\label{properdistance}
\end{eqnarray}

That is, we~measure proper time using the metric $g_{\mu \nu}$ but the space
geometry is determined by both metrics. In~this model massive particles do not
move on geodesics of a four-dimensional metric. Only~massless particles move
on a null geodesic of $\mathfrak{g}_{\mu \nu}$. Therefore,~Delta Gravity is not a
metric theory.

% -----------------------------------------------------------
% Section 2
% -----------------------------------------------------------
\section{\texorpdfstring{\boldmath{$T^{\mu\nu}$}}{Tun} and \texorpdfstring{$\tilde{T}^{\mu\nu}$}{TTun} for a Perfect Fluid}

The Energy-Stress Tensors for a Perfect Fluid in Delta Gravity are~\cite{DeltaGravityL} (assuming $c$ is the speed of light equal to 1):

\begin{equation}
T_{\mu\nu}\ =\ p(\rho)g_{\mu\nu}+\left(\rho+p(\rho)\right)U_{\mu}U_{\nu}
\end{equation}
\vspace{-27pt}

\begin{equation}
\begin{split}
\tilde{T}_{\mu \nu}\ = p(\rho)\tilde{g}_{\mu \nu}+\frac{\partial p}{\partial\rho}(\rho)\tilde{\rho}g_{\mu \nu}+\left(\tilde{\rho}+\frac{\partial p}{\partial\rho}(\rho)\tilde{\rho}\right)U_{\mu}U_{\nu}+\\
\left(\rho+p(\rho)\right)\left(\frac{1}{2}(U_{\nu}U^{\alpha}\tilde{g}_{\mu\alpha}+U_{\mu}U^{\alpha}\tilde{g}_{\nu\alpha})+U_{\mu}^{T}U_{\nu}+U_{\mu}U_{\nu}^{T}\right)
\end{split}
\end{equation}
where $U^\alpha U_\alpha ^T=0$. $p$ is the pressure, $\rho$ is the density and $U^\mu$ is the four-velocity. For~more details you can see~\cite{DeltaGravityL}.

\section{Friedman-Lemaître-Robertson-Walker (FLRW) Metric}

In this section, we~discuss the equations of motion for
the Universe described by the FLRW metric. We~use spatial curvature equal to
zero to agree with cosmological observations.

In the harmonic coordinate system, it is~\cite{DeltaGravityL}:
\begin{eqnarray}
\label{g FLRW}
g_{\mu \nu}dx^{\mu}dx^{\nu} &=& - c^2 dt^2 + R^2(t)\left(dx^2 + dy^2 + dz^2\right) \\
\label{gt FLRW}
\tilde{g}_{\mu \nu}dx^{\mu}dx^{\nu} &=& - 3F_a(t) c^2 dt^2 + F_a(t)R^2(t)\left(dx^2 + dy^2 + dz^2\right)
\end{eqnarray}
\vspace{-24pt}

\begin{equation}
\mathbf{g}_{\mu \nu}=g_{\mu\nu}+\ \tilde{g}_{\mu\nu}=-c^2(1+3\ F_a(t)) dt^2+(1+\ F_a(t))R^2(t) \left(dx^2+dy^2+dz^2\right)
\label{geodesics null}
\end{equation}

%----nuevo!!

Please note that $F_a(t)$ is an arbitrary function that remains after imposing homogeneity and isotropy of the space as well as the extended harmonic gauge $g^{\alpha \beta}\frac{1}{2}g^{\mu \lambda}\left(D_{\beta}\tilde{g}_{\lambda \alpha}+D_{\alpha}\tilde{g}_{\beta \lambda}-D_{\lambda}\tilde{g}_{\alpha \beta}\right) - \tilde{g}^{\alpha \beta}\Gamma^{\mu}_{\alpha \beta} = 0$.It is determined by solving the differential equations in \eqref{tilde Eq}.

%----.

These expressions represent an isotropic and homogeneous Universe. From~\eqref{properdistance} we already know that the proper time is measured only using the metric $g_{\mu \nu}$, but~the space geometry in FLRW coordinates is determined by the modified null geodesic, given by \eqref{geodesics null}, where both tensor fields, $g_{\mu \nu}$~and $\tilde{g}_{\mu \nu}$, are~needed.

%%%%%Equations Solution.%%%%%
\section{Delta Gravity Friedmann Equations}

The equations of state for matter and radiation are:

\[p_m(R)=0\]

\[p_r(R)=\frac{1}{3}\rho_r(R)\]

Then, from Equation~\eqref{Einst Eq} we obtain:
\begin{eqnarray}
\rho(R)=\rho_m(R)+\rho_r(R)\label{177}\\
p_r(R)=\frac{1}{3}\rho_r(R)\label{178}\\
t(Y)=\frac{2\sqrt{C}}{3H_0\sqrt{\Omega_{r0}}}\left(\sqrt{Y+C}(Y-2C)+2C^{3/2}\right)\label{CosmologicalTimeDeltaGravity}\\
Y(t)=\frac{R(t)}{R_0} \label{180}\\
R_0\equiv R(t=t_0)\equiv 1\\
\Omega_{r}\equiv \frac{\rho_r}{\rho_c}\\
\Omega_m\equiv \frac{\rho_m}{\rho_c}\\
\rho_c\equiv {\frac {3H^{2}}{8\pi G}}\label{CriticalDensity}\\
\Omega_{r0}+\Omega_{m0}\equiv 1\\
\Omega_{r0}=\frac{1}{1+\frac{1}{C}}
\end{eqnarray}
where $t_0$ is the age of the Universe (at the current time).
It is important to highlight that $t$ is the cosmic time, $R_0$ is the standard scale factor at the current time, $C\equiv \frac{\Omega_{r0}}{\Omega_{m0}}$, where $\Omega_{r0}$ and $\Omega_{m0}$ are the density energies normalized by the critical density at the current time, defined as the same as the Standard Cosmology. Furthermore,~we~have imposed that Universe must be flat ($k=0$), so~we require that $\Omega_r+\Omega_m\equiv 1$.\\

% Nuevo----

Using the Second Continuity Equation~\eqref{Conserv tilde T},where $\tilde{T}_{\mu \nu}$ is a new Energy-Momentum Tensor, two~new densities called $\tilde{\rho}_M$ (Delta Matter Density) and $\tilde{\rho}_R$ (Delta Radiation Density) associated with this new tensor are defined. When~we solve this equation, we~find
\begin{eqnarray}
\tilde{\rho}_M(Y)= -\frac{3\rho_{m0}}{2}\frac{F_a(Y)}{Y^3} \label{DeltaMatterDensity}\\
\tilde{\rho}_R(Y)=-2\rho_{r0}\frac{F_a(Y)}{Y^4} \label{DeltaRadiationDensity}
\end{eqnarray}

Using the Second Field Equation \eqref{tilde Eq} with the solutions \eqref{DeltaMatterDensity} and \eqref{DeltaRadiationDensity} we found (and redefining with respect to $Y$):

\begin{equation}
F_a(Y)=-\frac{L_2}{3}Y\sqrt{Y+C}
\label{Fa-redef}
\end{equation}

Then, writing Equations \eqref{DeltaMatterDensity} and \eqref{DeltaRadiationDensity} in terms of $L_2$ we have

\begin{eqnarray}
\tilde{\rho}_m(Y)= \left(\frac{L_2}{2}\right)\rho_{m0}\frac{\sqrt{Y+C}}{Y^2} \label{DeltaMatterDensityFinal}\\
\tilde{\rho}_r(Y)=\left(\frac{2L_2}{3}\right)\rho_{r0}\frac{\sqrt{Y+C}}{Y^3} \label{DeltaRadiationDensityFinal}
\end{eqnarray}

Thus, if~we know the $C$ and $L_2$ values, it~is possible to know the Delta Densities $\tilde{\rho}_m$ and $\tilde{\rho}_r$.

% ----

\subsection*{Relation between the Effective Scale Factor \texorpdfstring{${Y_{DG}}$}{YDG} and the Scale Factor \texorpdfstring{$Y$}{Y}}

The Effective Metric for the Universe is given by \eqref{geodesics null}. From~this expression, it~is possible to define the Effective Scale Factor as follows:

\begin{equation}
{R_{DG}}(t)=R(t)\sqrt{\frac{1+F_{a}(t)}{1+3 F_{a}(t)}}
\label{R_DG}
\end{equation}

Defining that $R(t_0)\equiv 1$, we~have that $R(t)=Y(t)$. Furthermore,~we~define the Effective Scale Factor (normalized):

\begin{equation}
{Y_{DG}}\equiv\frac{{R_{DG}}(t)}{{R_{DG}}(t_0)}=\frac{Y}{{R_{DG}}(t_0)}\sqrt{\frac{1-L_{2}\frac{Y}{3}\sqrt{Y+C}}{1-L_{2}Y\sqrt{Y+C}}}
\label{YtildeDef2}
\end{equation}

Please note that the denominator in Equation~\eqref{YtildeDef2} is equal to zero when $1=L_2 Y\sqrt{Y+C}$. Also~remember~that $C=\Omega_{r0}/\Omega_{m0}<<1$.  Furthermore,~we~have imposed that $\tilde{\rho}_m>0$ and $\tilde{\rho}_r>0$, then $L_2$ must be greater than $0$~\cite{DeltaGravityL}. Then~the valid range for $L_2$ is approximately $0\leq L_2\leq 1$.

$C$ must be positive, and~(hopefully) is a very small value because the radiation is clearly not dominant in comparison with matter. Then,~we~can analyze cases close to the standard accepted value: $\Omega_{r0}/\Omega_{m0}\sim 10^{-4}$.

\section{Useful Equations for Cosmology}

Here we present the equations that are useful to fit the SN Data and obtain cosmological parameters that are presented in the Results Section.

\subsection{Redshift Dependence}

The relation between the cosmological redshift and the scale factor is preserved in Delta Gravity:

\begin{equation}
{Y_{DG}}=\frac{1}{1+z}
\label{Ytilderedshift}
\end{equation}

It is important to take into account that the current time is given by $t_0\to Y(t_0)\to {Y_{DG}}(Y=1)=1$, where $Y_{DG}$ is normalized.

\subsection{Luminosity Distance}

The proof is the same as GR, because the main idea is based on the light traveling through a null geodesic described by the Effective Metric given by \eqref{geodesics null} in Delta Gravity~\cite{DG_DE}. Taking~into account that idea, we~can obtain the following expression:

\begin{equation}
d_L(z,L_2,C)=c\frac{(1+z)\sqrt{C}}{100\sqrt{h^2\Omega_{r0}}} \int_{Y(t_{1})}^{1} \frac{Y}{\sqrt{Y+C}}\frac{dY}{{Y_{DG}}(t)}
\label{LuminosityDistanceDG}
\end{equation}

Notice that $Y=1$ today. To~solve $Y(t_1)$ at a given redshift $z$, we~need to solve \eqref{YtildeDef2} and \eqref{Ytilderedshift} numerically. Furthermore,~the~integrand contains ${Y_{DG}}(t)$ that can be expressed in function of $Y$ in \eqref{YtildeDef2}. Do~not confuse $c$ (speed of light) with $C$, a~free parameter to be fitted by SN Data.

% Nuevo ---
The parameter $h^2 \Omega_{r0}$ can be obtained from the CMB. The~CMB Spectrum can be described by a Black Body Spectrum, where the energy density of photons is given by

\[\rho_{\gamma 0}=aT^4\]

From statistical mechanics, we~know the neutrinos are related by~\cite{Planck2015}:

\[\rho_{\nu 0} = 3\frac{7}{8}\left(\frac{4}{11}\right)^{4/3}\rho_{\gamma 0}\]

Then,

\begin{equation}
\Omega_{r 0}h^2 = \Omega_{\gamma}h^2 + \Omega_{\nu}h^2
\label{Omegarh2}
\end{equation}

Equation \eqref{Omegarh2} is a value that only depends on the temperature of the Black Body Spectrum of the CMB. So we can add this value as a known Cosmological Parameter.
%Newly added information, please confirm.

% ---Agree

Thus, we~only need to know the values $C$ and $L_2$. Take~into account that it is impossible to know the value of $\Omega_{r0}$ without any other information.

\subsection{Distance Modulus}

The distance modulus is the difference between the apparent magnitude $m$ and the absolute magnitude $M$ of an astronomical object. Knowing~this we can estimate the distance $d$ to the object, provided that we know the value of the absolute magnitude $M$.

\begin{equation}
{\displaystyle \mu =m-M=5\log _{10}\left({\frac {d}{10\,\mathrm {pc} }}\right)}
\label{DistanceModulus}
\end{equation}

\subsection{Effective Scale Factor}

The ``size'' of the Universe in Delta Gravity is given by ${Y_{DG}}(t)$, while in GR this is given by $a(t)$. Every~cosmological parameter that in the GR theory was built up from the standard scale factor $a(t)$, in~Delta Gravity will be built from ${Y_{DG}}(t)$. This~value is equal to $1$ at the current time, because it is the $R_{DG}$ normalized by $R_{DG}(Y=1)$.

\subsection{Hubble Parameter} \label{hubble}

In Delta Gravity we will define the Hubble Parameter as follows:

\begin{equation}
H^{DG}(t)\equiv\frac{{\dot{R}_{DG}}(t)}{{R_{DG}}(t)}
\end{equation}

Therefore, the~Hubble Parameter is given by:

\begin{equation}
H^{DG}(t)=\frac{\frac{d {R_{DG}}}{dY}\left(\frac{dt}{dY}\right)^{-1}}{{R_{DG}}}
\label{HDG}
\end{equation}

Notice that all the Delta Gravity parameters are written as function of $Y$.

\subsection{Deceleration Parameter}

In Delta Gravity we will define the deceleration parameter as follows:

\begin{equation}
q^{DG}(t)=-\frac{{\ddot{R}_{DG}}{R_{DG}}}{{\dot{R}_{DG}}^2}
\end{equation}

Then,

\begin{equation}
q^{DG}(t)=-\frac{\frac{d }{dY}\left(\frac{d {R_{DG}}}{dY}\left(\frac{dt}{dY}\right)^{-1}\right)\left(\frac{dt}{dY}\right)^{-1}{R_{DG}}}{\left(\frac{d {R_{DG}}}{dY}\left(\frac{dt}{dY}\right)^{-1}\right)^2}
\label{DGqparameter}
\end{equation}

\section{Fitting the SN Data}

We are interested in the viability of Delta Gravity as a real Alternative Cosmology Theory that could explain the accelerating Universe without $\Lambda$, then it is natural to check if this model fits the SN~Data.

\subsection{SN Data}

To analyze this, we used the most updated Type Ia Supernovae Catalog. We~obtained the Data from Scolnic~\cite{Scolnic2018}. We~only needed the distance modulus $\mu$ and the redshift $z$ to the SN-Ia to fit the model using the Luminosity Distance $d_L$ predicted from the theory.

The SN-Ia are very useful in cosmology~\cite{Riess1998} because they can be used as standard candles and allow to fit the $\Lambda$CDM model finding out free parameters such as $\Omega_\Lambda$. We~are interested in doing this in Delta Gravity. The~main characteristic of the SN-Ia that makes them so useful is that they have a very standardized absolute magnitude close to $-19$~\cite{Riess2016, Mabs-19.05,Mabs-19.05BetoulePeroEnLibro,Mabs-19.25,Mabs-19.26}.

% Nuevo ----

From the observations we only know the apparent magnitude and the redshift for each SN-Ia. Thus,~we~have the option to use a standardized absolute magnitude obtained by an independent method that does not involve $\Lambda$CDM model, or~any other assumptions.

To fit the SN-Ia Data, we~will use $M_V=-19.23\pm 0.05$~\cite{Riess2016}. The~value was calculated using 210 SN-Ia Data from~\cite{Riess2016}. This~value is independent from the model since it was calculated by building the distance ladder starting from local Cepheids measured by parallax and using them to calibrate the distance to Cepheids hosted in near galaxies (by Period-Luminosity relations) that are also SN-Ia host. Riess~et al. calculated the $M_V$ and the $H_0$ local value, and~they did not use any particular cosmological model. Keep~in mind that the value of $M_V$ found by Riess et al. is an intrinsic property of SN-Ia and that is the reason they are used as standard candles.

We used 1048 SN-Ia Data in~\cite{Scolnic2018}\footnote{Scolnic's Data are available at \url{https://archive.stsci.edu/hlsps/ps1cosmo/scolnic/}.}. All~the SN-Ia are spectroscopically confirmed. In~this paper, we~have used the full set of SN-Ia presented in~\cite{Scolnic2018}. They~present a set of spectroscopically confirmed PS1 SN-Ia and combine this sample with spectroscopically confirmed SN-Ia from CfA1-4, CSP, PS1, SDSS, SNLS and HST SN surveys.

At~\cite{Scolnic2018} they used the SN Data to try to obtain a better estimation of the DE state equation. They~define the distance modulus as follows:

\begin{equation}
\mu\equiv m_B-M+\alpha x_1-\beta c+\Delta_M+\Delta_B
\label{muDefScolnic}
\end{equation}
where $\mu$ is the distance modulus, $\Delta_M$ is a distance correction based on the host-galaxy mass of the SN and $\Delta_B$ is a distance correction based on predicted biases from simulations. Furthermore,~$\alpha$ is the coefficient of the relation between luminosity and stretch, $\beta$ is the coefficient of the relation between luminosity and color and $M_V$ is the absolute B-band magnitude of a fiducial SN-Ia with
$x_1 = 0$ and $c = 0$~\cite{Scolnic2018}.

In this work we are not interested in the specific corrections to observational magnitudes of SN-Ia. We~only take the values extracted from~\cite{Scolnic2018} to analyze the Delta Gravity model. The~SN Data are the redshift $z_i$ and $(\mu+M)_i$ with the respective errors.

% -----

\subsection{Delta Gravity Equations}

We need to establish a relation between redshift and the apparent magnitude for the SN-Ia:

\begin{equation}
[\mu+M]-M=5\log _{10}\left({\frac {d_L(z,C,L_2)}{10\,\mathrm {pc} }}\right)
\end{equation}
where $d_L(z,L_2,C)$ is given by \eqref{LuminosityDistanceDG} and $[\mu+M]$ are the SN-Ia Data given at~\cite{Scolnic2018}.

In this expression we have as free parameters: $C$ and $L_2$ to be found by fitting the model to the points $(z_i, [\mu+M]_i)$.

\subsection{GR Equations}

For GR we use the following expression

\begin{equation}
[\mu+M]-M=5\log _{10}\left({\frac {d_L(z,H_0,\Omega_{m0})}{10\,\mathrm {pc} }}\right)
\end{equation}
where $d_L(z,H_0,\Omega_{m0})$ is given by:

\begin{equation}
d_L(z,H_0,\Omega_{m0})=\frac{c(1+z)}{H_0}\int_{\frac{1}{1+z}}^{1} \frac{du}{\sqrt{(1-\Omega_{m0})u^4+\Omega_{m0}u}}
\label{RGLuminosityDistance}
\end{equation}
and $[\mu+M]$ are the SN Data given at~\cite{Scolnic2018}. Remember~that we are always working on a flat Universe, and~in GR standard model the $\Omega_{r0}$ is negligible. We~have the same degrees of freedom as Delta Gravity.

Please note that we are including DE as $\Omega_{\Lambda 0}\equiv \Omega_{\Lambda}\equiv 1-\Omega_{m 0}$ in GR.

\subsection{MCMC Method}

To fit the SN-Ia Data to GR and Delta Gravity models, we~used Markov Chain Monte Carlo (MCMC). This~routine was implemented in Python 3.6 using PyMC2.\footnote{\url{https://pymc-devs.github.io/pymc/.}}

Basically, MCMC consists on fitting a model, characterizing its posterior distribution. It~is based on Bayesian Statistics. We~used the Metropolis-Hastings algorithm.

% Nuevo ----
We used a Bayesian approach because it allows us to know the posterior probability distribution for every parameter of the model~\cite{BayesianBook,PiattellaNotes}. Furthermore,~it~is possible to identify dependencies between the fitted parameters using MCMC, which it is not possible using another method such as the least-square used in~\cite{DeltaGravityL}.
% ----------

Initially we propose initial distributions for the parameters that we want to fix, and~then PyMC2 will give us the posterior probability distribution for these parameters.

We want to find the best fitted parameters for Delta Gravity and GR models. These~parameters will be $C,L_2$ for Delta Gravity and $H_0,\Omega_M$ for GR.

\section{Results and Analysis}

We present the results for Delta Gravity and GR fitted Data, and~with these values we obtain different cosmological parameters. We~divide the results into two fits: { Delta Gravity Fit} and { GR Fit}.

\subsection{Fitted Curves}

%----nuevo!

\begin{figure}[H]
\centering
\includegraphics[width=12cm,keepaspectratio]{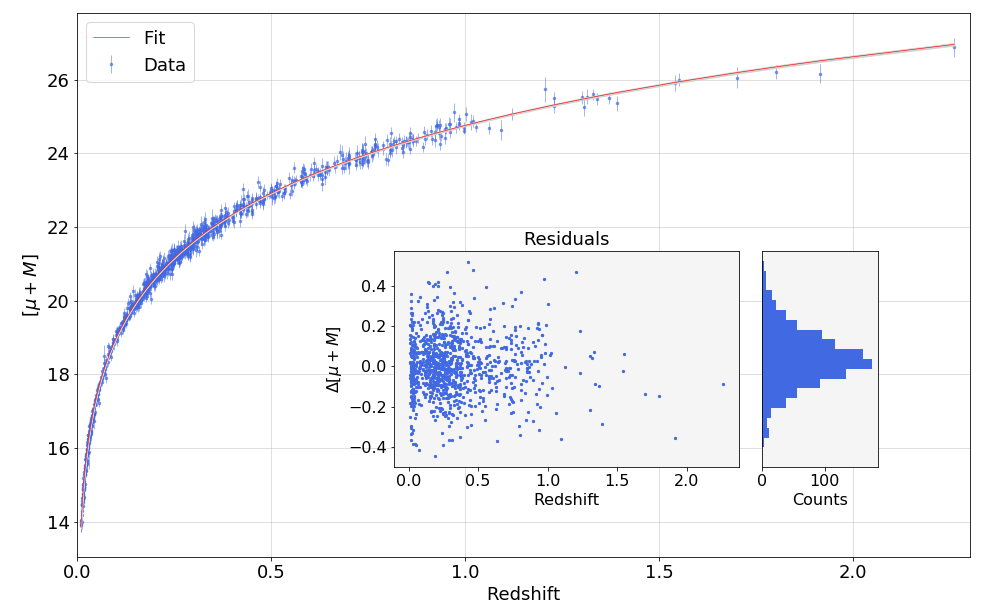}
\caption{Fitted curve for Delta Gravity model assuming $M_V=-19.23$. On~the right corner, the~residual plot for the fitted Data.}
\label{Fit_DG}
\end{figure}

\begin{figure}[H]
\centering
\includegraphics[width=12cm,keepaspectratio]{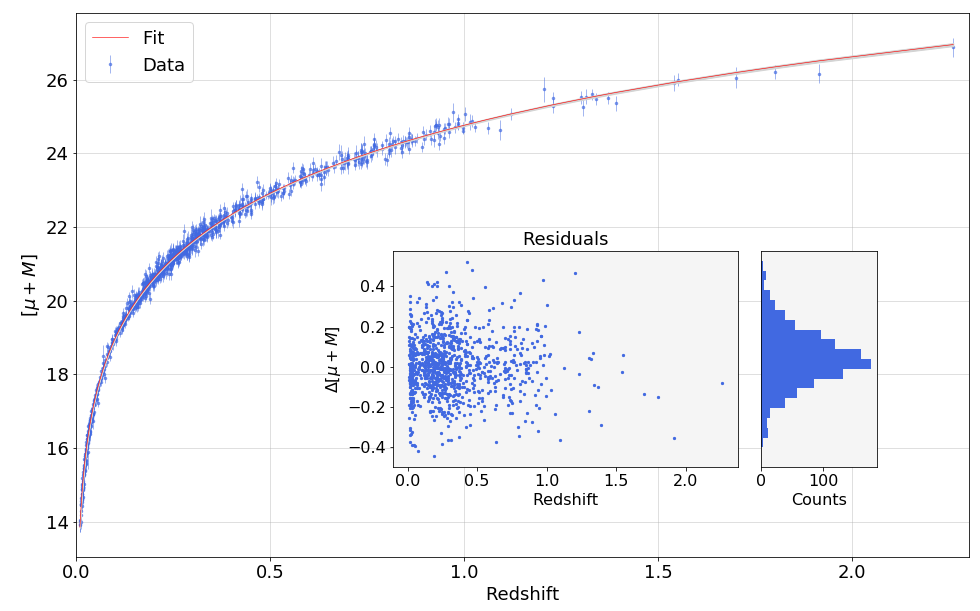}
\caption{Fitted curve for GR standard model assuming $M_V=-19.23$. On~the right corner, the~residual plot for the fitted Data. }
\label{Fit_GR}
\end{figure}

As we see in Figures \ref{Fit_DG} and \ref{Fit_GR}, both models describe very well the $m_B$ vs. $z$ SN-Ia Data. It~is important to note that, while in GR frame $\Lambda \neq 0$ is needed to find this well-behaved curve, in~Delta Gravity $\Lambda$ is not needed to fit the SN-Ia Data. Essentially,~Delta Gravity predicts the same behavior, but~the accelerating Universe appears explained without the need to include $\Lambda$, or~anything like ``Dark~Energy''.

In Table \ref{tableMfixed}, we~present the coefficients of determination ($r^2$) and residual sum of squares (RSS) for both fitted models:

\begin{table}[H]
\caption{Statistical parameters.}
\centering
\begin{tabular}{ccc}
\toprule
\textbf{Model} & \boldmath{$r^2$} & \boldmath{$RSS$} \\
\midrule
Delta Gravity & 0.99709 & 21.39 \\
GR & 0.99708 & 21.44\\
\bottomrule
\end{tabular}
\label{tableMfixed}
\end{table}

Both coefficients of determination are very good, and the RSS are similar for both cases.

The fitted parameters for GR and Delta Gravity models are shown in Tables \ref{tableMfixedDG} and \ref{tableMfixedGR} respectively.

\begin{table}[H]
\caption{Fitted parameters using MCMC for Delta Gravity.}
\centering
\begin{tabular}{ccc}
\toprule
\textbf{Delta Gravity} & \textbf{Value} & \textbf{Error} \\
\midrule
$L_2$ & 0.455& 0.008 \\
$C$ & 0.000169 & 0.000003 \\
\bottomrule
\end{tabular}
\label{tableMfixedDG}
\end{table}
\unskip

\begin{table}[H]
\caption{Fitted parameters using MCMC for GR.}
\centering
\begin{tabular}{ccc}
\toprule
\boldmath{$M_V$} \textbf{Fixed GR Model} & \textbf{Value} & \textbf{Error} \\
\midrule
$\Omega_{m0}$ & 0.28 & 0.01 \\
$h^2$ & 0.549& 0.004\\
\bottomrule
\end{tabular}
\label{tableMfixedGR}
\end{table}
\unskip

\unskip

\begin{figure}[H]
\begin{subfigure}{.5\textwidth}
\centering
\includegraphics[width=\linewidth]{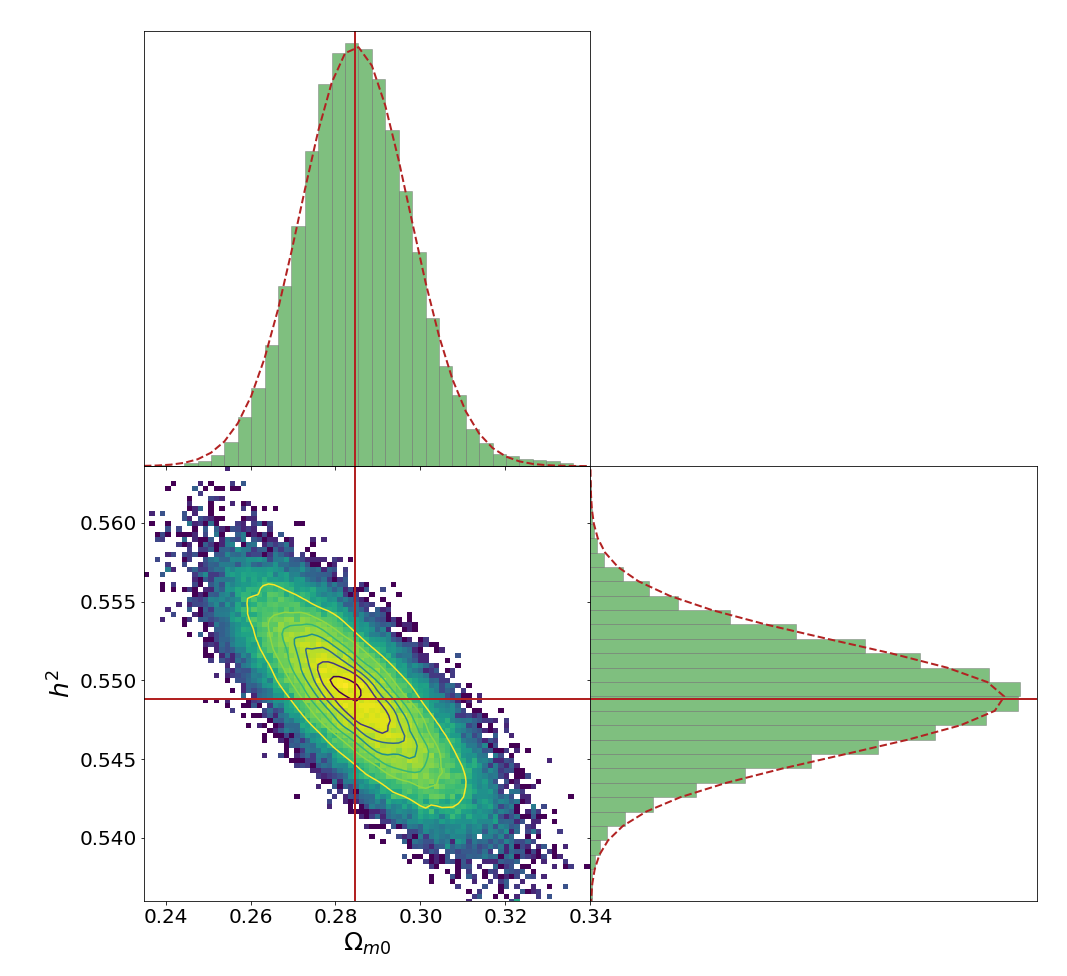}
\caption{}
\label{FigurePPmapsGR}
\end{subfigure}%
\ \ \
\begin{subfigure}{.5\textwidth}
\centering
\includegraphics[width=\linewidth]{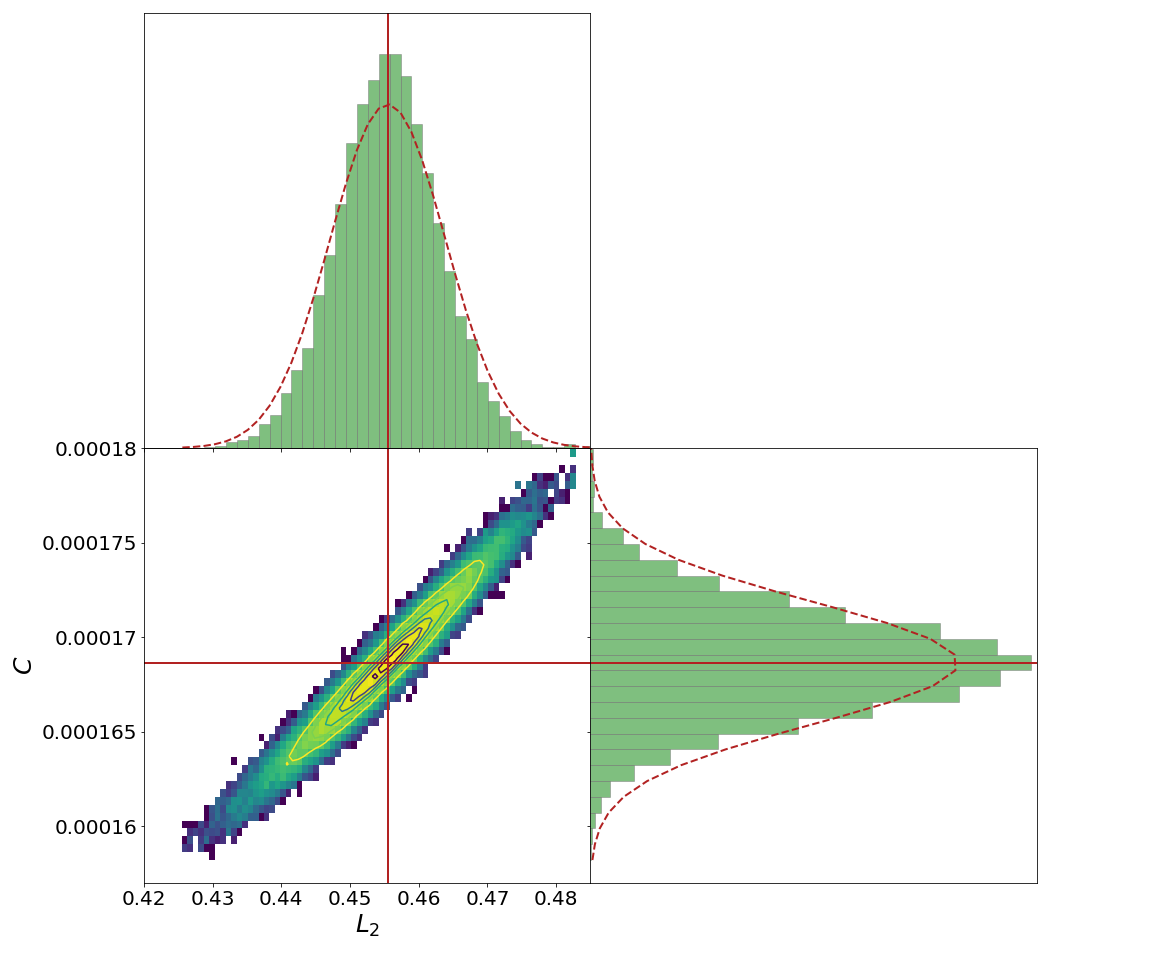}
\caption{}
\label{FigurePPmapsDG}
\end{subfigure}
\caption{Posterior probability density plots obtained from MCMC for GR and Delta Gravity models. (\textbf{a})~Posterior probability density maps with for GR. Combination~for $\Omega_{m0}$ and $h^2$. (\textbf{b})~Posterior probability density maps for Delta Gravity. Combination for $L_2$ and $C$.}
\label{FigurePPmaps} 
\end{figure}

Furthermore, we~present the posterior probability density maps for GR and Delta Gravity in Figure \ref{FigurePPmaps}.

Please note that for both plots in Figure \ref{FigurePPmaps} the distributions are well defined, and~for each parameter we obtain a Gaussian-like distribution. For~both models, the~combination of parameters constrained a region in the 2D-density plot. The~fitted values for both models converged very well.

% NUevo ----
\subsection{Convergence Tests}

\begin{figure}[H]
\begin{subfigure}{.5\textwidth}
\centering
\includegraphics[width=\linewidth]{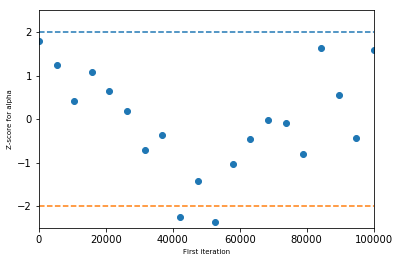}
\caption{}
\label{geweke_GR}
\end{subfigure}%
\begin{subfigure}{.5\textwidth}
\centering
\includegraphics[width=\linewidth]{geweke_DG.png}
\caption{}
\label{geweke_DG}
\end{subfigure}
\caption{Convergence of values for GR and Delta Gravity. (\textbf{a})~Evolution of $z$-scores with steps in GR. (\textbf{b})~Evolution of $z$-scores with steps in Delta Gravity.}
\label{geweke}
\end{figure}

We applied two convergence tests for MCMC analysis. The~first is known as Geweke~\cite{Geweke}. This~is a time-series approach that compares the mean and variance of segments from the beginning and end of a single chain.
This method calculates values named $z$-scores (theoretically distributed as standard normal variates). If the chain has converged, the majority of points should fall within 2 standard deviations of zero\footnote{\url{https://media.readthedocs.org/pdf/pymcmc/latest/pymcmc.pdf}.}. The~plots are shown in Figure \ref{geweke}.

In both plots it is possible to observe that the most part of the $z$-scores fall within $2\sigma$, so~the method is convergent for both models based on the Geweke criterion.

Another convergence test is the Gelman-Rubin statistic~\cite{Gelman}.

The Gelman-Rubin diagnostic uses an analysis of variance approach to assess convergence. This~diagnostic uses multiple chains to check for lack of convergence, and~is based on the notion that if multiple chains have converged, by~definition they should appear very similar to one another; if not, one~or more of the chains has failed to converge (see PyMC 2 documentation).

In practice, we~look for values of $\hat{R}$ close to one because this is the indicator that shows~convergence.

\begin{figure}[H]
\centering
\includegraphics[width=10cm,height=10cm]
{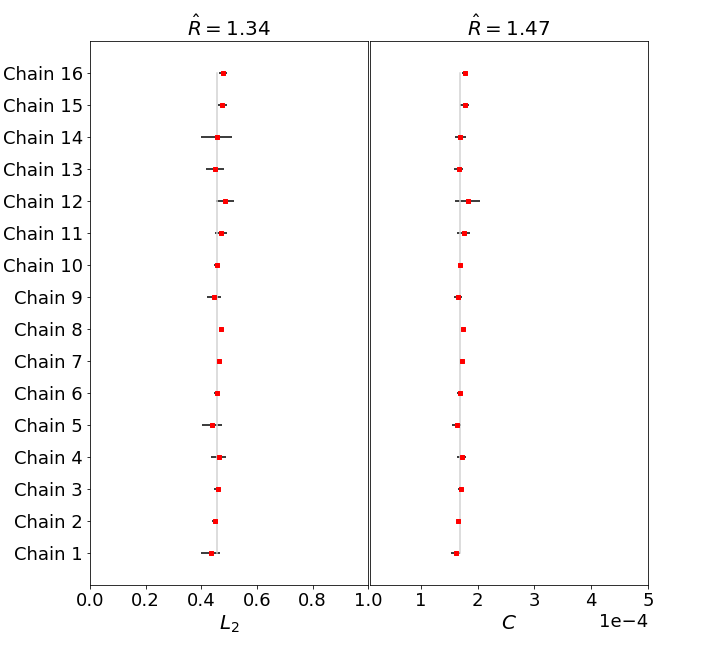}
\caption{Gelman-Rubin test for Delta Gravity model assuming $M_V=-19.23$. The~Gelman-Rubin test was run with 16 different chains, all~with different $L_2$ and $C$ priors. The~$\hat{R}$ coefficient (Gelman-Rubin coefficient) was calculated for each parameter. }
\label{Gelman_Rubin_DG.png}
\end{figure}

\begin{figure}[H]
\begin{subfigure}{.5\textwidth}
\centering
\includegraphics[width=\linewidth]{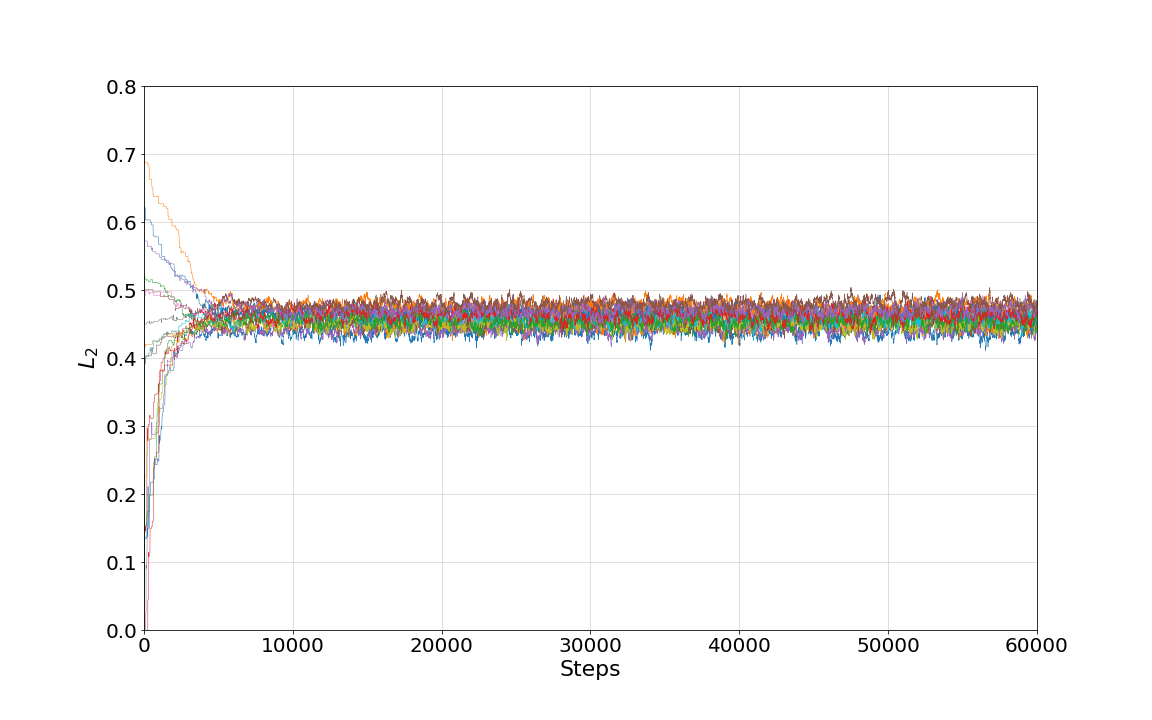}
\caption{}
\label{Multichains_Convergence_L2.png}
\end{subfigure}%
\ \ \
\begin{subfigure}{.5\textwidth}
\centering
\includegraphics[width=\linewidth]{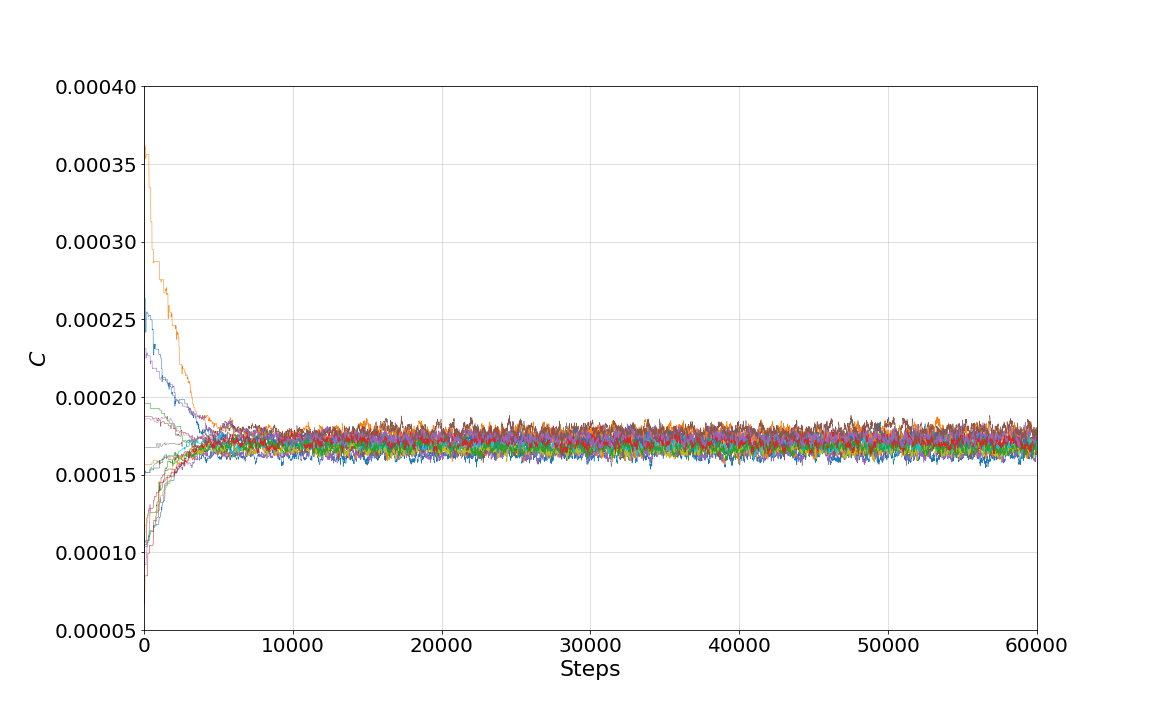}
\caption{}
\label{Multichains_Convergence_C.png}
\end{subfigure}
\caption{Gelman-Rubin test for Delta Gravity model. There~are 16 chains with different priors. (\textbf{a})~All the chains converge to a $L_2\approx 0.455$. (\textbf{b})~All the chains converge to a $C\approx 0.000169$.}
\label{Multichains_Convergence}
\end{figure}

We ran 16 chains for Delta Gravity model. Figure~\ref{Gelman_Rubin_DG.png} shows the $L_2$ and $C$ predicted values for every chain of the Monte Carlo simulation. Figure~\ref{Multichains_Convergence}a,b shows the convergence of $L_2$ and $C$. All~the chains converge to a similar value assuming different priors. These final values predicted for every chain can be visualized in Figure \ref{Gelman_Rubin_DG.png}. From~all these chains, is~clear that the Delta Gravity MCMC analysis is convergent for the two free parameters.

\unskip

% ------

\subsection{Cosmic Time and Redshift}

By using Equation~\eqref{CosmologicalTimeDeltaGravity} we obtain the Cosmic Time in Delta Gravity, where the redshift is obtained by numerical solution from Equation~\eqref{Ytilderedshift}.

Meanwhile for GR model, we~obtained the cosmic time from the integration of the first Friedmann equation and solving $t(\Omega_{m0},H_0)$. Here~we have included $\Omega_{\Lambda}=1-\Omega_{m0}$ and we did $\Omega_{k}$ ($k=0$) and $\Omega_{r0}=0$. The~integral for the first Friedmann equation can be analytically solved:
\begin{equation}
t=\int_0^a\frac{1}{\sqrt[]{\frac{\Omega_{m0}}{x}+(1-\Omega_{m0})x^2}}dx = \frac{2}{3\sqrt[]{1-\Omega_{m0}}} \ln\left(\frac{\sqrt[]{-\Omega_{m0}a^3+\Omega_{m0}+a^3}+\sqrt[]{1-\Omega_{m0}}a^{3/2}}{\sqrt[]{\Omega_{m0}}}\right)
\label{CosmicTimeGR}
\end{equation}
where $t$ in \eqref{CosmicTimeGR} is the cosmic time for GR.

%nuevo ---
We plot the results in Figure \ref{FigTimeredshift}:
\begin{figure}[H]
\centering
\includegraphics[width=12cm,keepaspectratio]{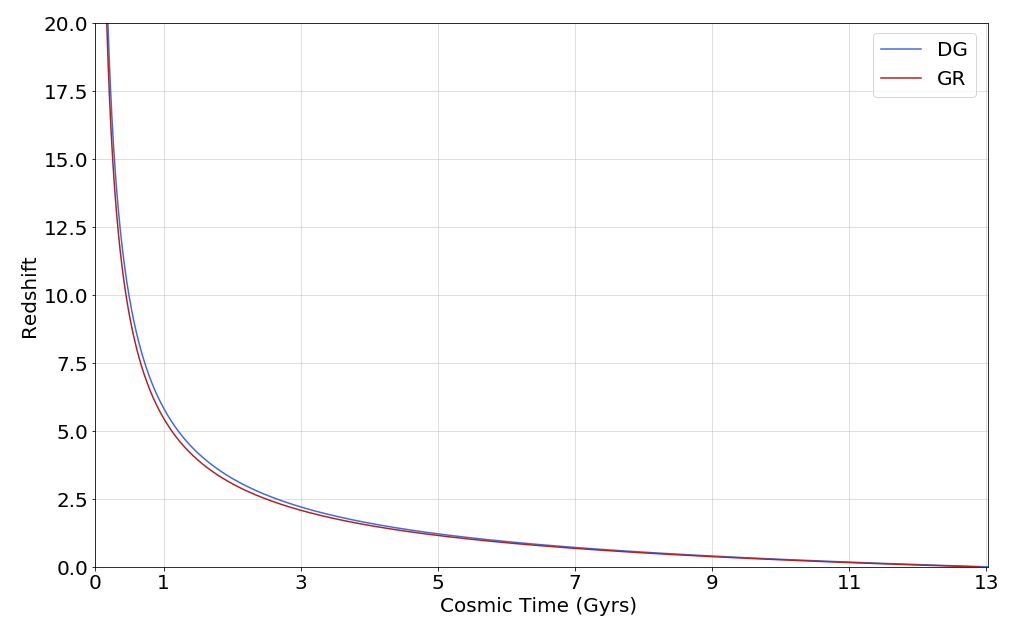}
\caption{cosmic time for GR and Delta Gravity.}
\label{FigTimeredshift}
\end{figure}
% ----

The behavior of cosmic time dependence with redshift for both models is very similar.

\subsection{Hubble Parameter and \texorpdfstring{$H_0$}{H0}}

% nuevo ----
With the fitted parameters found by MCMC for GR and Delta Gravity, we~can find $H(t)$ and $H_0$. Note~the superscript for GR as $^{GR}$ and Delta Gravity as $^{DG}$. For~GR $H_0$ is easily obtained from the $h^2$ fitted ($H_0=100h$). $H^{GR}(t)$ can be obtained using the first Friedmann equation

\begin{equation}
H^2 = \left(\frac{\dot{a}}{a}\right)^2 = \frac{8 \pi G}{3}\left(\frac{\rho_{m0}}{a^3}+\frac{\rho_{r0}}{a^4}+\rho_{\Lambda 0}\right)
\label{FirstFriedmannEquation}
\end{equation}

Taking into account that $\Omega_{m0}+\Omega_{r0}+\Omega_{\Lambda 0}=1$, $\Omega_{r0}\approx 0$, and~$\rho_{c0}=\frac{3 H_0^2}{8\pi G}$, where $\Omega_{X_i0}=\frac{\rho_{X_i0}}{\rho _{c0}}$ for every $X_i$ component in the Universe, we~obtain

\begin{equation}
H^2 = H_0^2\left(\frac{\Omega_{m0}}{a^3}+(1-\Omega_{m0})\right)
\label{FirstFriedmannEquation2}
\end{equation}

\begin{figure}[H]
\centering
\includegraphics[width=11cm,keepaspectratio]{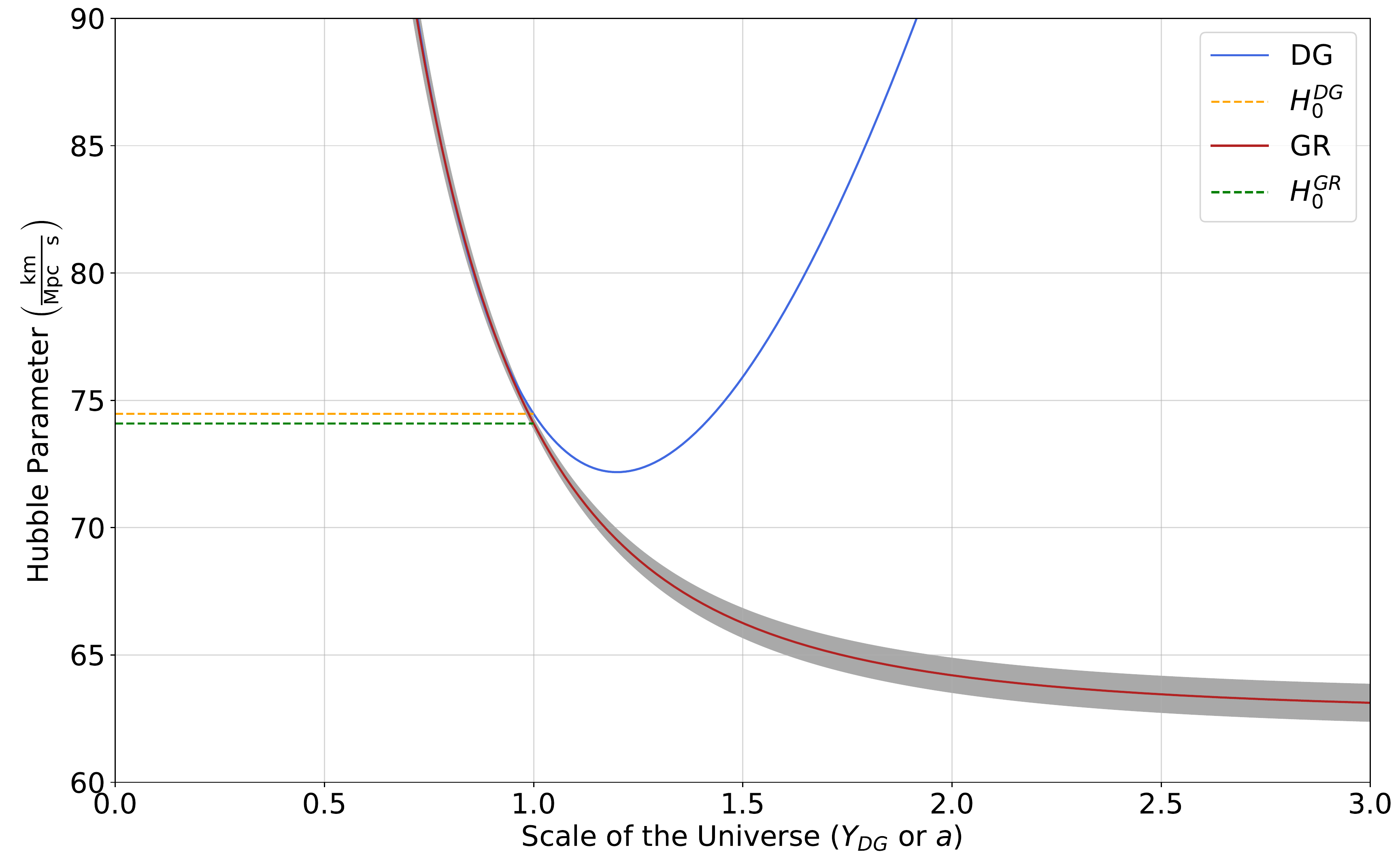}
\caption{Hubble Parameter for Delta Gravity and GR fitted models assuming $M_V = -19.23$.} %Please add figure citation in the main text. ADDED
\label{HvsScaleFactor}
\end{figure}

With \eqref{FirstFriedmannEquation2}, we~obtain $H^{GR}(t)$ and using \eqref{HDG} we obtain $H^{DG}(t)$, Figure \ref{HvsScaleFactor}. For~the actual time we evaluate $H^{GR}$ at $a=1$ and for Delta Gravity we evaluate $H^{DG}$ at $Y_{DG}=1$ obtaining the Hubble constant $H_0^{GR}$ and $H_0^{DG}$.

% ----

We present the results for both models and we compare these values with previous measurements in Table \ref{tableH0compared}.

\begin{table}[H]
\centering
\caption{$H_0$ values found by MCMC with SN-Ia Data, assuming $M_V = -19.23$. Furthermore, we~tabulate Planck \cite{Planck2018} and Riess \cite{Riess2018} $H_0$ values.}  %Please add table citation in the main text. ADDED
\label{tableH0compared}
\begin{tabular}{ccc}
\toprule
\textbf{Model} & \boldmath{$H_0$} \textbf{( km/(s Mpc)} ) & \textbf{Error} \\
\midrule
Planck 2018~\cite{Planck2018} & 67.36 & 0.54 \\
Riess 2018~\textsuperscript{a}~\cite{Riess2018} & 73.52 & 1.62 \\
Riess 2018~\textsuperscript{b}~\cite{Riess2018} & 73.83 & 1.48 \\
GR & 74.08 & 0.24 \\
Delta Gravity & 74.47 & 1.63\\
\bottomrule
\end{tabular}

\begin{tabular}{c}
\multicolumn{1}{p{\linewidth}}{\footnotesize \centering{\textsuperscript{a} The calibration was made including the new MW parallaxes from \emph{HST} and \emph{Gaia}; \textsuperscript{b} The calibration was made considering the external constrains on the parallax offset based on Red~Giants. }}
\end{tabular}

\end{table}
\unskip

\subsection{Age of the Universe}

The age of the Universe in Delta Gravity is calculated using \eqref{CosmologicalTimeDeltaGravity}. $t(Y)$ only depends on $C$ and not on $L_2$.
In GR we calculate the age of the Universe using \eqref{CosmicTimeGR}.

With these expressions, we~can compare the behavior between cosmic time and the scale factor in GR (or the effective scale factor in Delta Gravity).

\begin{figure}[H]
\centering
\includegraphics[width=11cm,keepaspectratio]{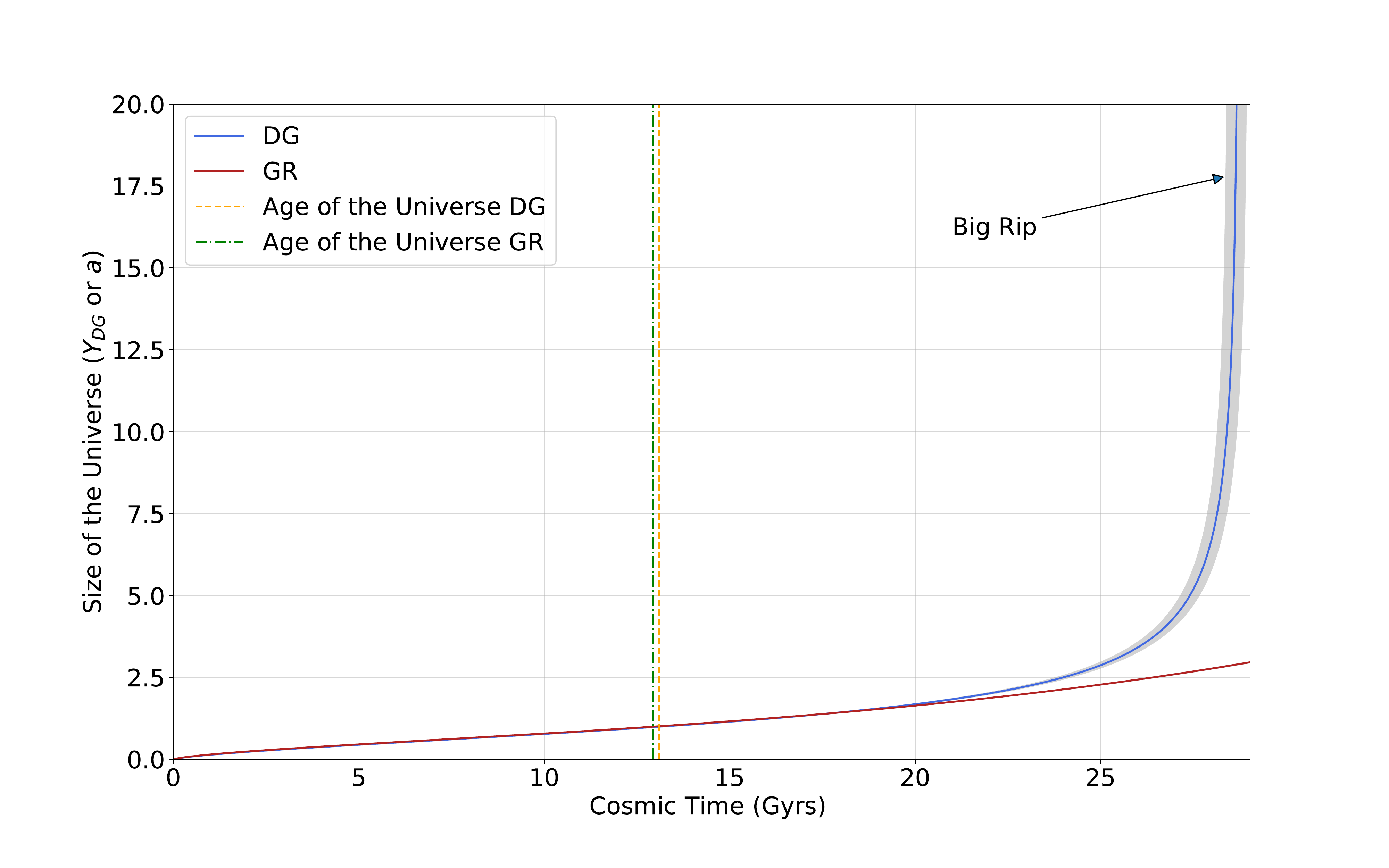}
\caption{The size of the Universe vs. age of the Universe. In~the Delta Gravity model, the~size of the Universe $Y_{DG}$ depends on cosmic time $t$ and on $C$. The~blue line indicates the effective scale factor in Delta Gravity. The~gray zone shows the error associated with $Y_{DG}$. For~GR, the~scale factor $a$ depends on cosmic time $t$ and on $\Omega_{m0}$. The~red line indicates the scale factor evolution in GR. The~gray zone shows the error associated with $a$ (these are tiny).}
\label{DGEffectiveScaleFactorWithTime}
\end{figure}

In Figure \ref{DGEffectiveScaleFactorWithTime}, it~is possible to see the evolution for $Y_{DG}(t)$ in time. At~$t=28.75$ Gyr, $Y_{DG}$ goes to infinity, and~the Universe ends with a Big Rip, then, in~this model the Universe has an end (in time). Also,~we~see the dependence between the scale factor $a$ and cosmic time $t$. The~Universe has no end (in~time) in GR.

\subsection{Deceleration Parameter \texorpdfstring{$q_0$}{q0}}

For Delta Gravity, we~used Equation~\eqref{DGqparameter}. For~today, we~evaluate $a=1$ for GR, and~$Y_{DG}=1$ for Delta Gravity.

%nuevo ----

\begin{figure}[H]
\begin{subfigure}{.5\textwidth}
\centering
\includegraphics[width=\linewidth]{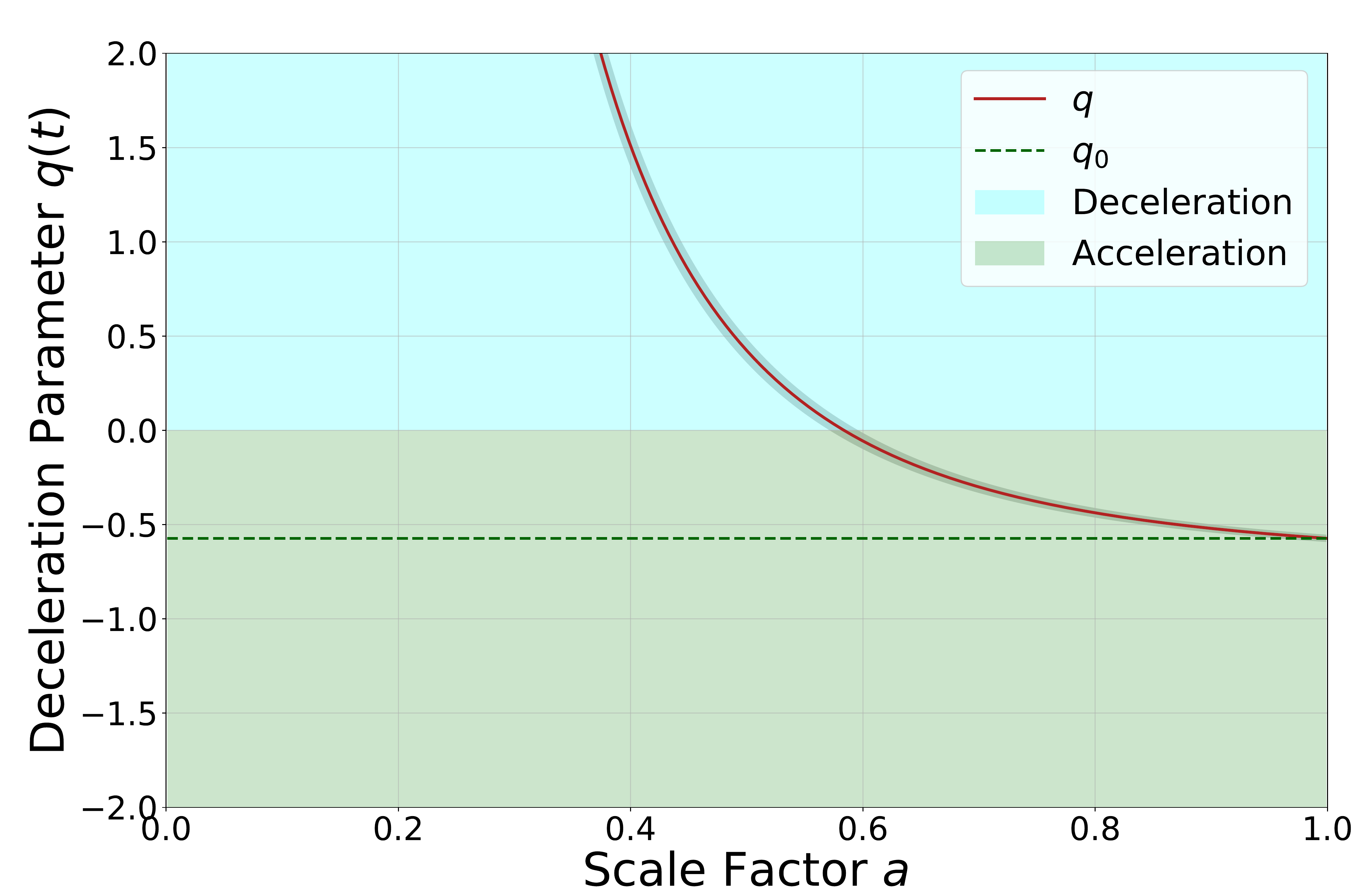}
\caption{}
\label{FigureConvergenceGR}
\end{subfigure}
\ \ \
\begin{subfigure}{.5\textwidth}
\centering
\includegraphics[width=\linewidth]{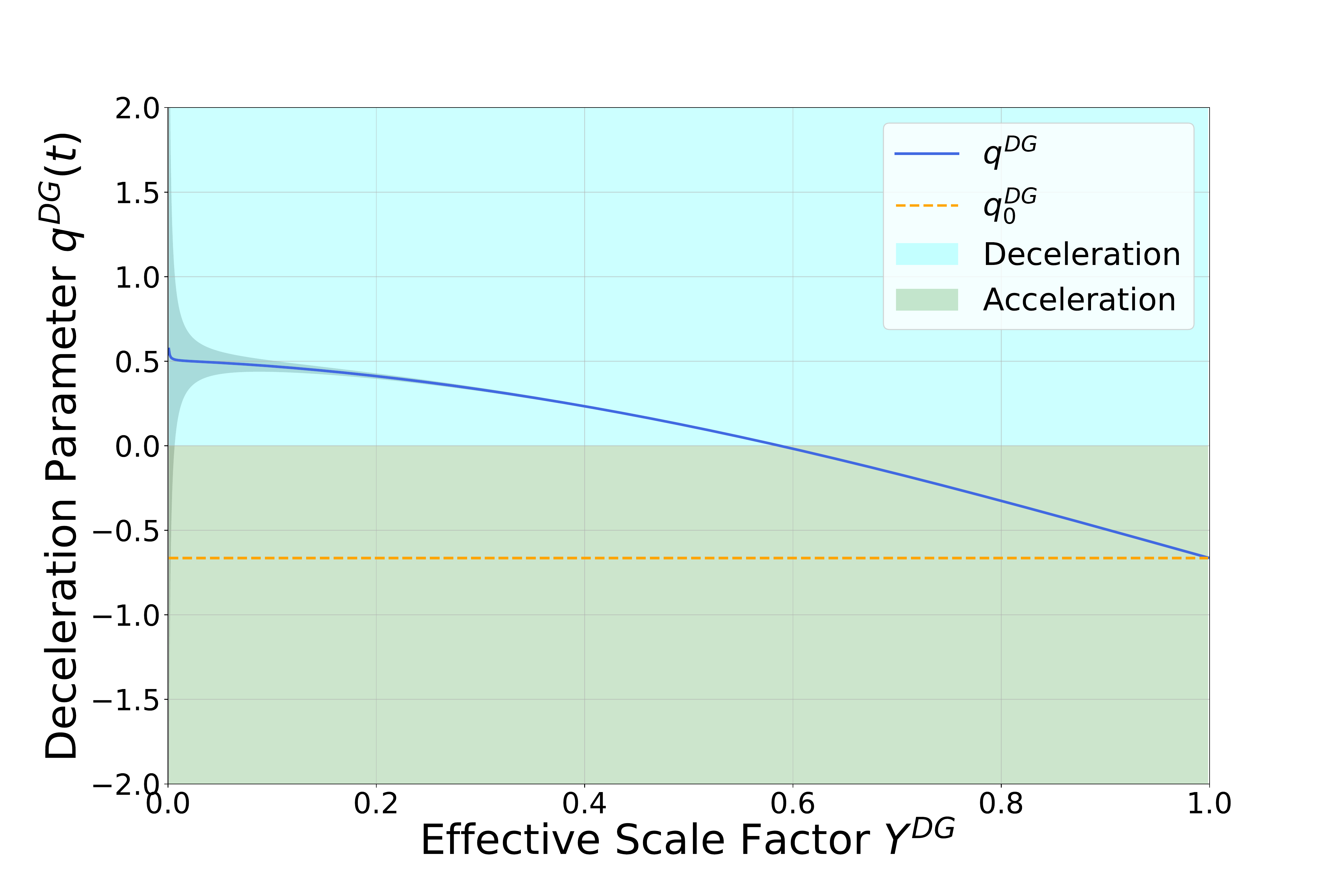}
\caption{}
\label{FigureConvergenceDG}
\end{subfigure}
\caption{Deceleration parameter for both models. (\textbf{a})~Evolution of deceleration parameter in GR. (\textbf{b})~Evolution of deceleration parameter in Delta Gravity.}
\label{qParameterPlots}
\end{figure}

In Figure \ref{qParameterPlots}, we~can see the evolution in time for both GR and Delta Gravity models.

% ------
We tabulate the deceleration parameter for both models in Table \ref{tableq0compared}.

\begin{table}[H]
\caption{$q_0$ values found by MCMC with SN-Ia Data, assuming $M_V = -19.23$.}
\label{tableq0compared}
\centering
\begin{tabular}{ccc}
\toprule
\textbf{Model} & \boldmath{$q_0$} & \textbf{Error} \\
\midrule
DG & $-$0.664 & 0.002 \\
GR & $-$0.57 & 0.02\\
\bottomrule
\end{tabular}
\end{table}

In both models $q_0<0$, then the Universe is accelerating.

% Nuevo ----

\subsection{Relation with Delta Components}

In Delta Gravity we are interested in determining the Delta composition of the Universe. Using~Equations \eqref{DeltaMatterDensityFinal} and \eqref{DeltaRadiationDensityFinal}, we~can obtain the densities for Delta Matter and Delta Radiation with the $C$ and $L_2$ fitted values.
\begin{eqnarray}
\tilde{\rho}_{m0}&= 0.22777\rho_{m0}&= 0.22773 \rho_{c0}
\label{DeltaMResult}\\
\tilde{\rho}_{r0}&=0.68330 \rho_{r0}&=0.000115 \rho_{c0}
\label{DeltaRResult}
\end{eqnarray}

In the expressions \eqref{DeltaRResult} and \eqref{DeltaMResult}, we~have obtained the current values for Delta Densities.

\begin{figure}[H]
\centering
\includegraphics[width=\linewidth]{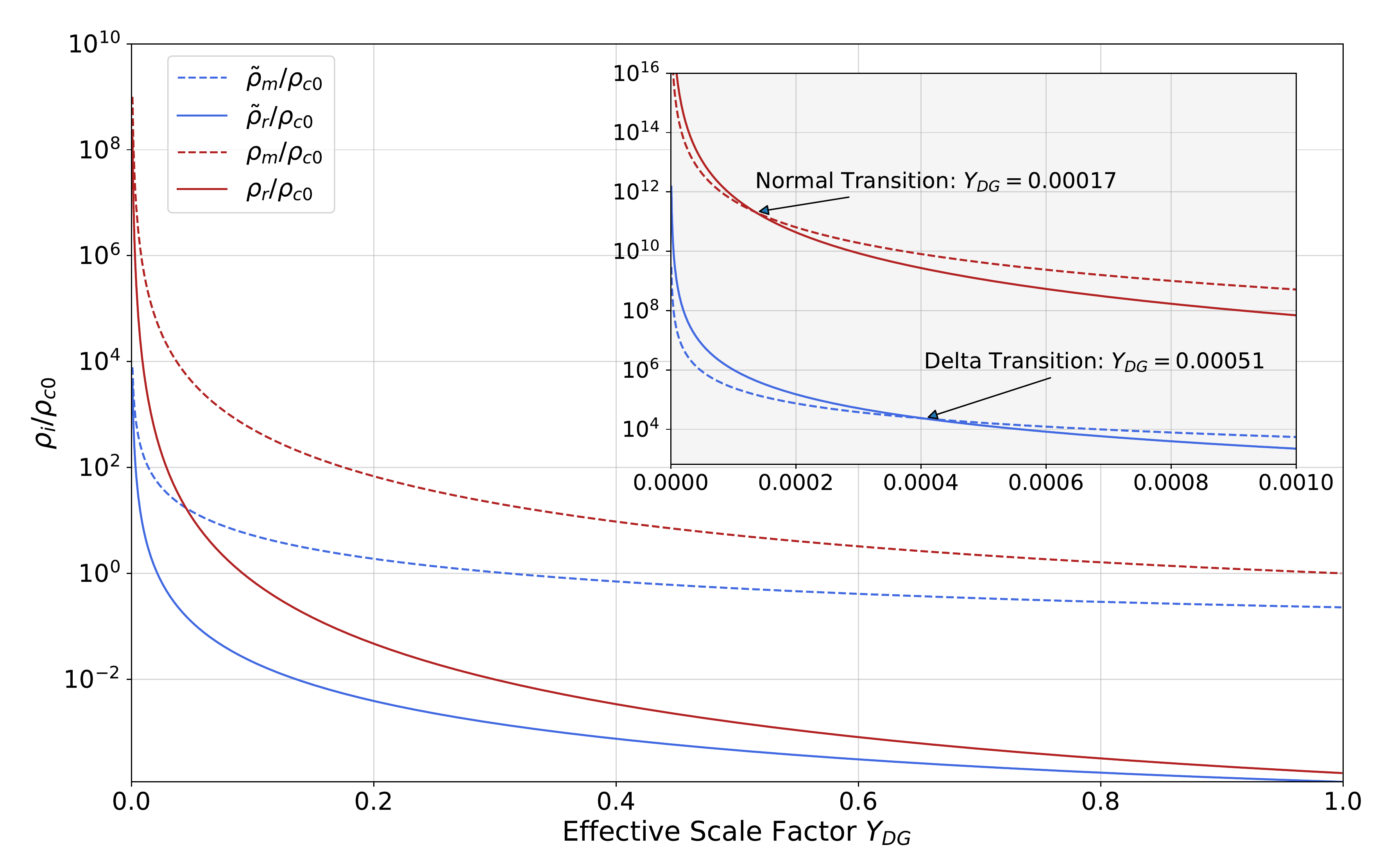}
\caption{Temporal evolution of density components for Delta Gravity. The~vertical axis is normalized by critical density at current time $\rho_{c0}$. On~the top right corner, there is a zoom in very close to $Y_{DG}=0$ showing the transition between Delta Matter and Delta Radiation (Delta components), and~the transition between matter and radiation (common components). In~general, the~Common Density is higher than the Delta Density.}
\label{Delta_Components}
\end{figure}

The Common Components are dominant compared with Delta components. Matter~is always dominant compared with radiation (in both cases). See~Figure \ref{Delta_Components}.

Please note that the four components diverge (in density) at the beginning of the Universe, and~the Delta Components show a ``constant-like'' behavior for $Y_{DG}>0.4$. (Specially Delta Matter that is clearly dominant compared to the Delta Radiation).

In both the Common Components and Delta Components, there is a transition between matter and radiation that is indicated in the zoom in included in Figure \ref{Delta_Components}. These~transitions occur at very early stage of the Universe. Both~transitions are indicated in Figure \ref{Delta_Components}.

It is important to remember that in Delta Gravity we do not know the $\rho_{c0}$, but~we know the densities of each component in units of $\rho_{c0}$, because they are given by $C$ and $L_2$ fitted values from SN~Data.
\unskip

% ---------

\subsection{Importance of \texorpdfstring{$L_2$}{L2} and \texorpdfstring{$C$}{C}}

\begin{figure}[H]
\begin{subfigure}{.5\textwidth}
\centering
\includegraphics[width=\linewidth]{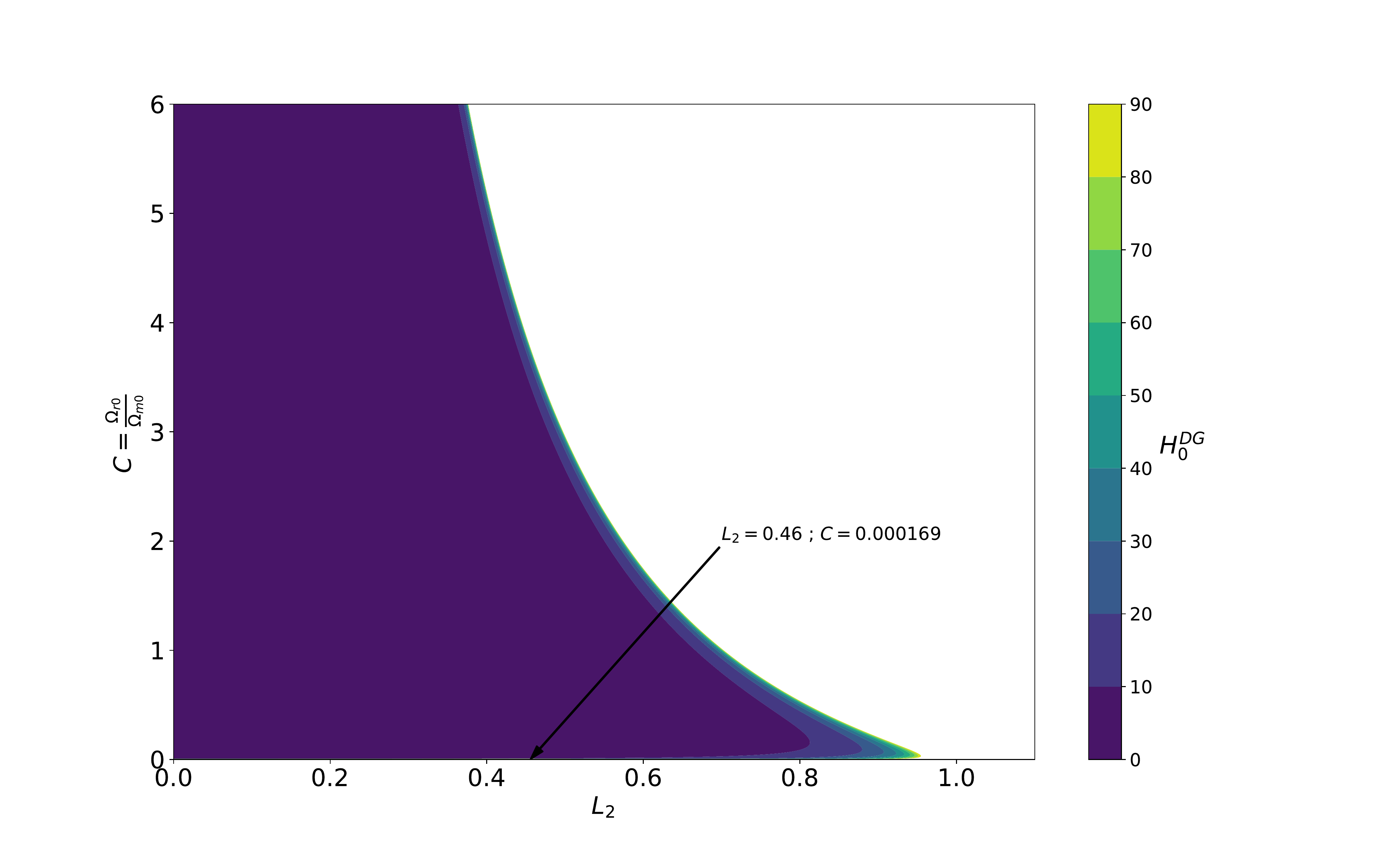}
\caption{}
\label{left_H0_contoursplot}
\end{subfigure}%
\ \ \
\begin{subfigure}{.5\textwidth}
\centering
\includegraphics[width=\linewidth]{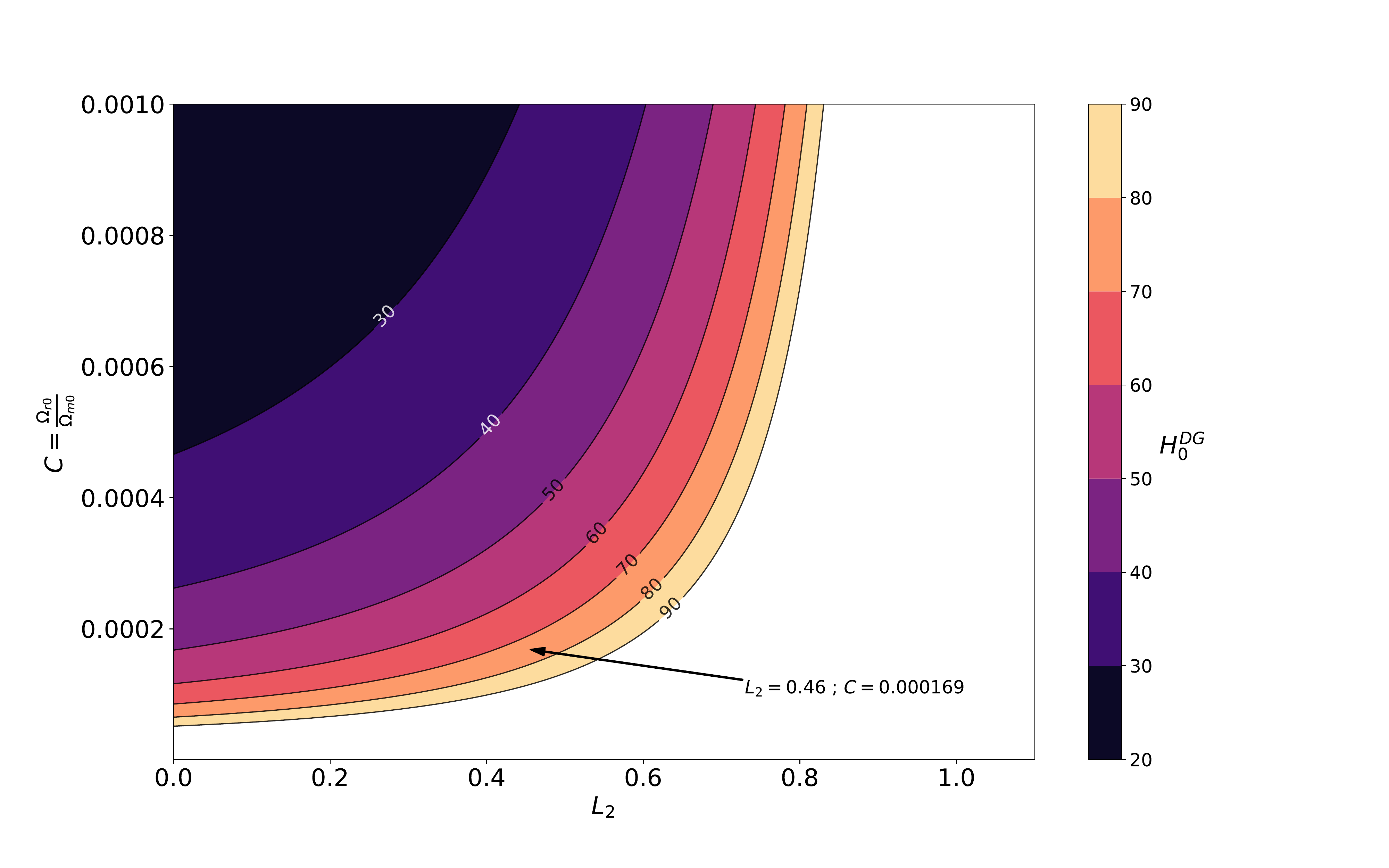}
\caption{}
\label{right_H0_contoursplot}
\end{subfigure}
\caption{$H_0^{DG}$ for a different combination of $L_2$ and $C$ values. The~fitted values found by MCMC analysis is indicated in the Figure. (\textbf{a})~$C$ values go from $0$ to $6$ to explore various Universes, even a Universe wholly dominated by radiation. (\textbf{b})~The $C$ values are bounded to very little values, nearly~close to the $C$ fitted value obtained by MCMC.}
\label{H0_contoursplot}
\end{figure}

\begin{figure}[H]
\begin{subfigure}{.5\textwidth}
\centering
\includegraphics[width=\linewidth]{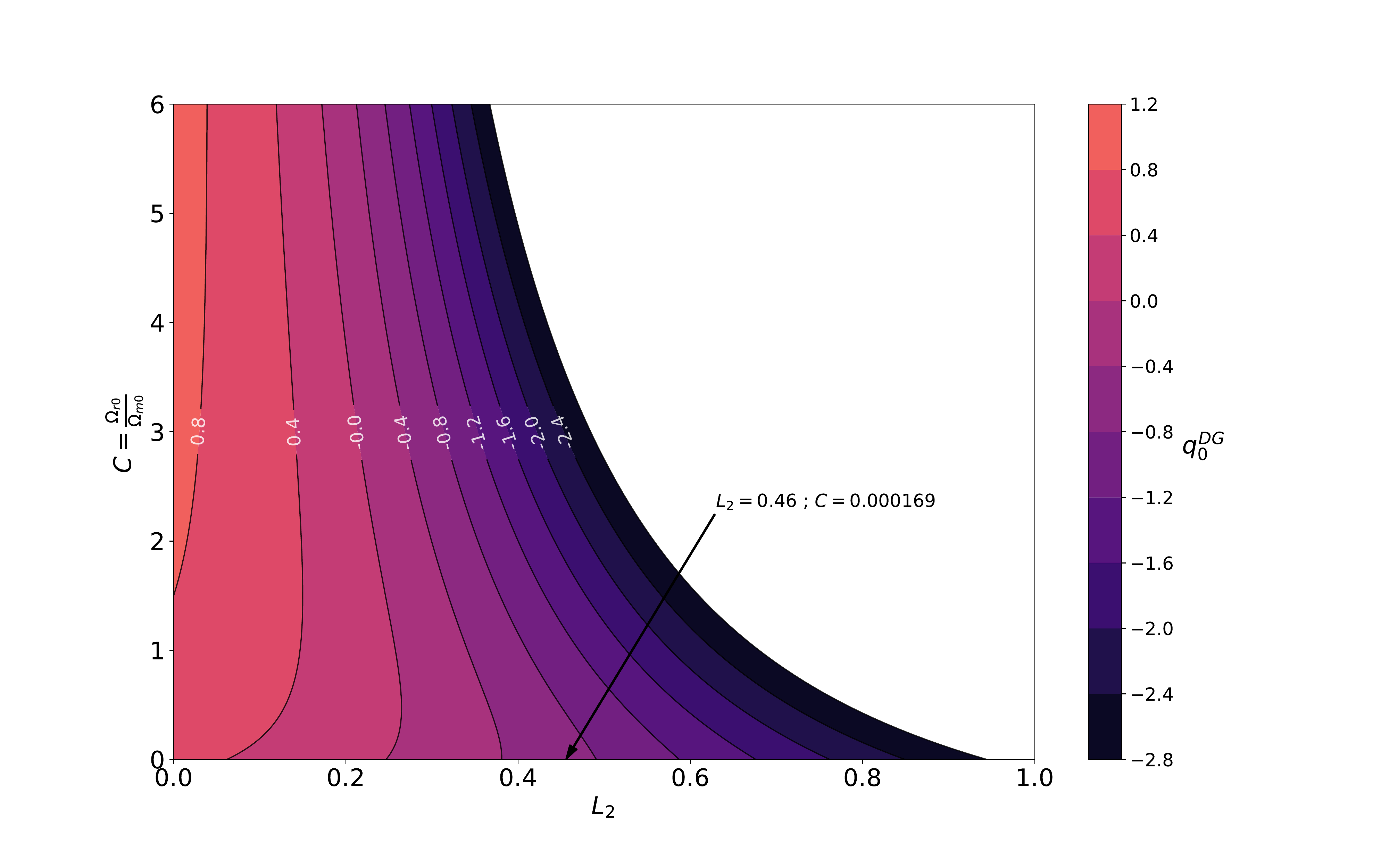}
\caption{}
\label{left_q0_contoursplot}
\end{subfigure}%
\ \ \
\begin{subfigure}{.5\textwidth}
\centering
\includegraphics[width=\linewidth]{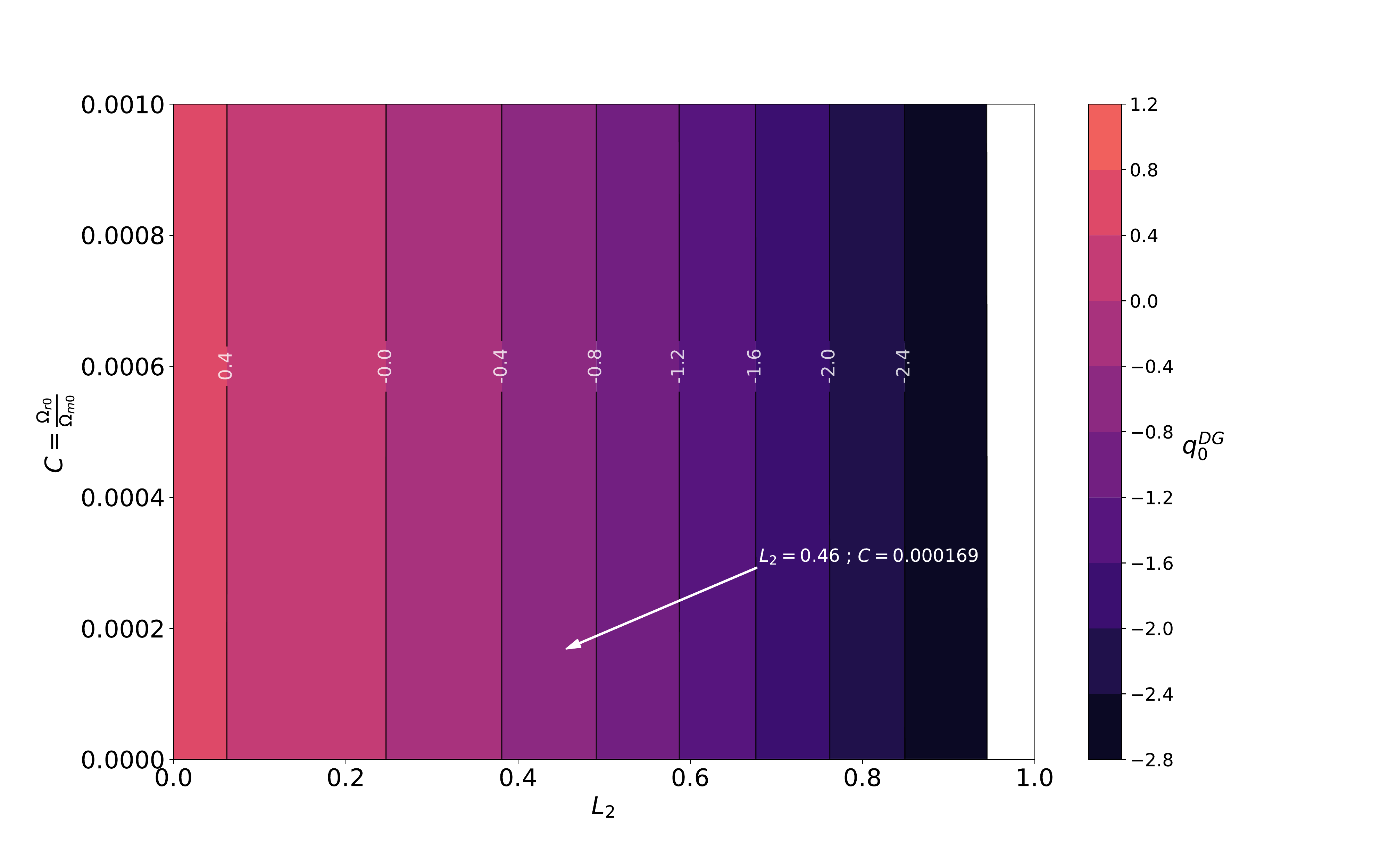}
\caption{}
\label{right_q0_contoursplot}
\end{subfigure}
\caption{$q_0^{DG}$ for different combination of $L_2$ and $C$ values. The~fitted values found by MCMC analysis is indicated in the Figure. (\textbf{a})~$C$ values go from $0$ to $6$ to explore various Universes, even a Universe wholly dominated by radiation. (\textbf{b})~The $C$ values are bounded to very little values, nearly~close to the $C$ fitted value obtained by MCMC.}
\label{q0_contoursplot}
\end{figure}

To understand the role that $L_2$ and $C$ are playing in the Delta Gravity model, we~need to plot some cosmological parameters in function of both coefficients. We~are interested in analyzing the accelerating expansion of the Universe in function of these two parameters, so~we plotted $H_0^{DG}$ in Figure \ref{H0_contoursplot} and $q_0^{DG}$ in Figure \ref{q0_contoursplot}.

% Nuevo -----

In Figure \ref{H0_contoursplot}, we~can see there is a big zone prohibited, because the results become complex values at certain level of the equations. The~only allowed values are colored. Note~that in Figure \ref{H0_contoursplot}a almost all the allowed $H_0^{DG}$ values are close to $0$. Only~the contour of the colored area shows $H_0^{DG}\neq 0$. The~Figure \ref{H0_contoursplot}b is the same as the left one, but~with a big zoom in close the fitted values obtained from MCMC analysis. These~range of $C$ and $L_2$ are reasonable to make an analysis. Note~that $H_0^{DG}$ has a strong dependence of $C$ and $L_2$ values.

Remember that $L_2$ has only sense between values $0$ and $1$, because we only want to allow positive Delta Densities and, from Equation \eqref{YtildeDef2}, the~denominator could be equal to $0$.

% ---------

The Figure \ref{q0_contoursplot} is very interesting because it shows the dependence of the current value of acceleration of the Universe expressed by the deceleration parameter $q_0^{DG}$. If~we examine the parameters zone close to the fitted values in the Figure \ref{q0_contoursplot}b, we~can note that the acceleration of the Universe only depends on the value of $L_2$. This~is a very important result from the Delta Gravity model. The~accelerating Universe is given by the $L_2$ parameter. This~parameter appears naturally like an integration constant from the differential equations when we solved the field equations for Delta Gravity model. Then,~in~this model, and~exploring the closest area to the Universe with a little amount of radiation compared to matter, we~found that a higher $L_2$ value, higher the acceleration of the Universe (current age): $q_0^{DG}$ becomes more negative when $L_2 \to 1$ independently of $C$.

\section{$M$ Free}

For completeness, we~want to mention that we also did the MCMC analysis for $M$ free in both GR and Delta Gravity.

\begin{figure}[H]
\begin{subfigure}{.5\textwidth}
\centering
\includegraphics[width=\linewidth]{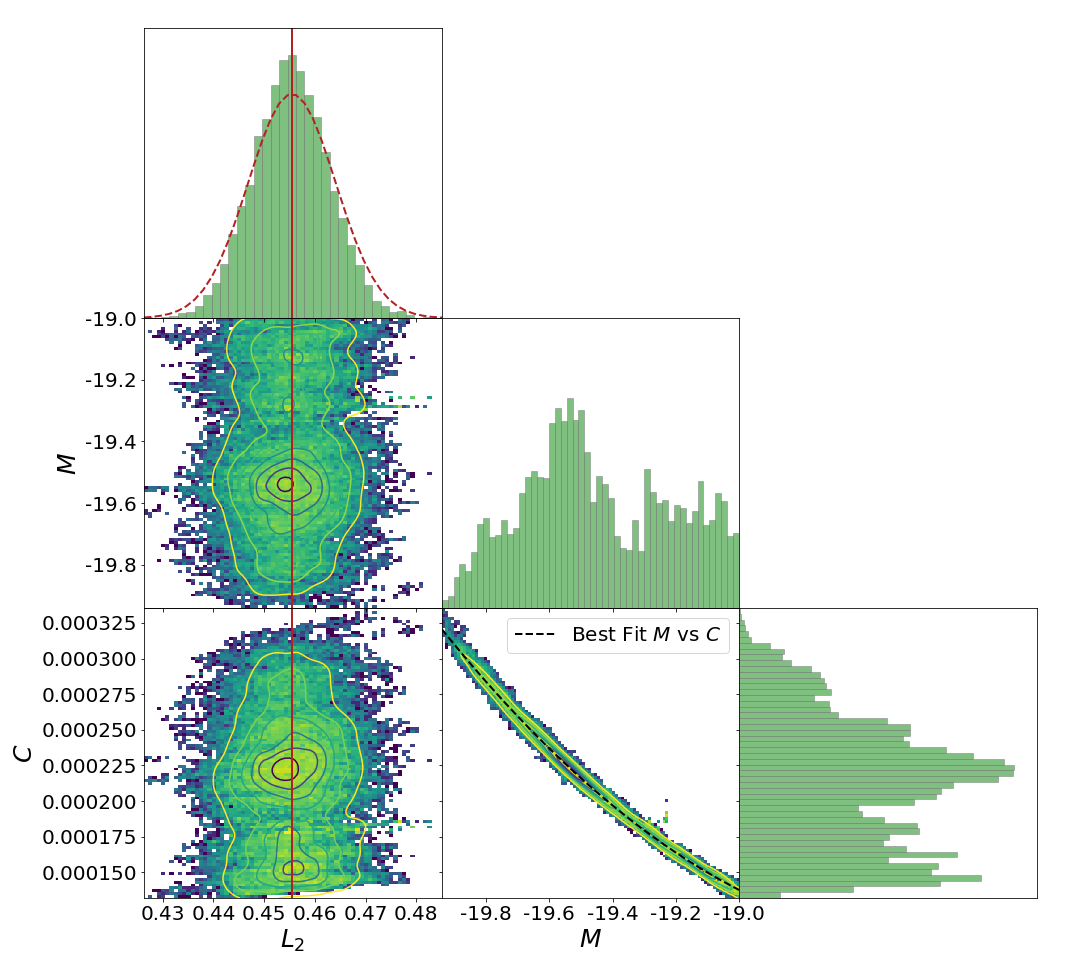}
\caption{}
\end{subfigure}%
\begin{subfigure}{.5\textwidth}
\centering
\includegraphics[width=\linewidth]{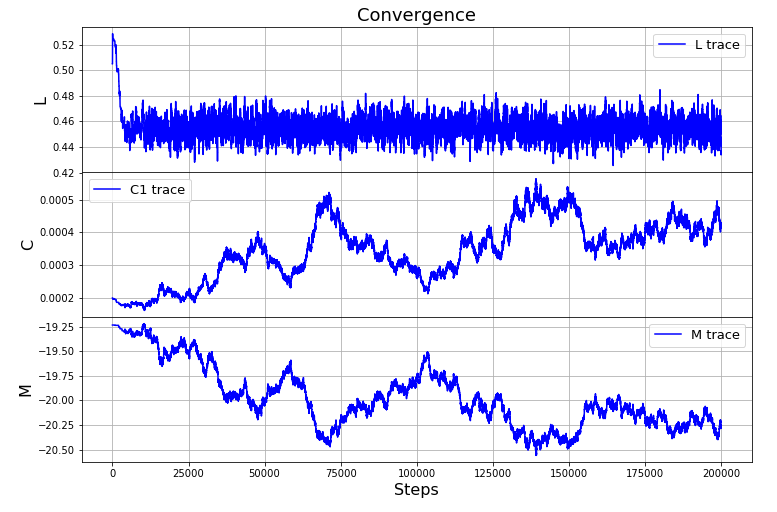}
\caption{}
\end{subfigure}
\caption{MCMC analysis assuming $M$ as a free parameter in Delta Gravity. (\textbf{a})~Posterior probabilities densities. (\textbf{b})~Evolution of values with steps. }
\label{DGMfreeplots}
\end{figure}

From the MCMC analysis, we~obtain a non-convergent result. In~Delta Gravity model, the~$C$ and $M$ parameters are dependent, but~$L_2$ is independent.
This can be visualized in Figure \ref{DGMfreeplots}.

The dependence for Delta Gravity parameters can be fitted by a second order polynomial, as~shown in Figure \ref{DGMfreeplots}:

\begin{equation}
C = 8.59\times10^{-5}M^2 + 3.15 \times 10^{-3} M + 2.9 \times 10^{-2}
\label{polinomioDG_M_C}
\end{equation}

If we use $M=-19.23\pm 0.05$~\cite{Riess2016}, we~fix $C$ which agrees
with the results of the previous sections.

\begin{figure}[H]
\begin{subfigure}{.5\textwidth}
\centering
\includegraphics[width=\linewidth]{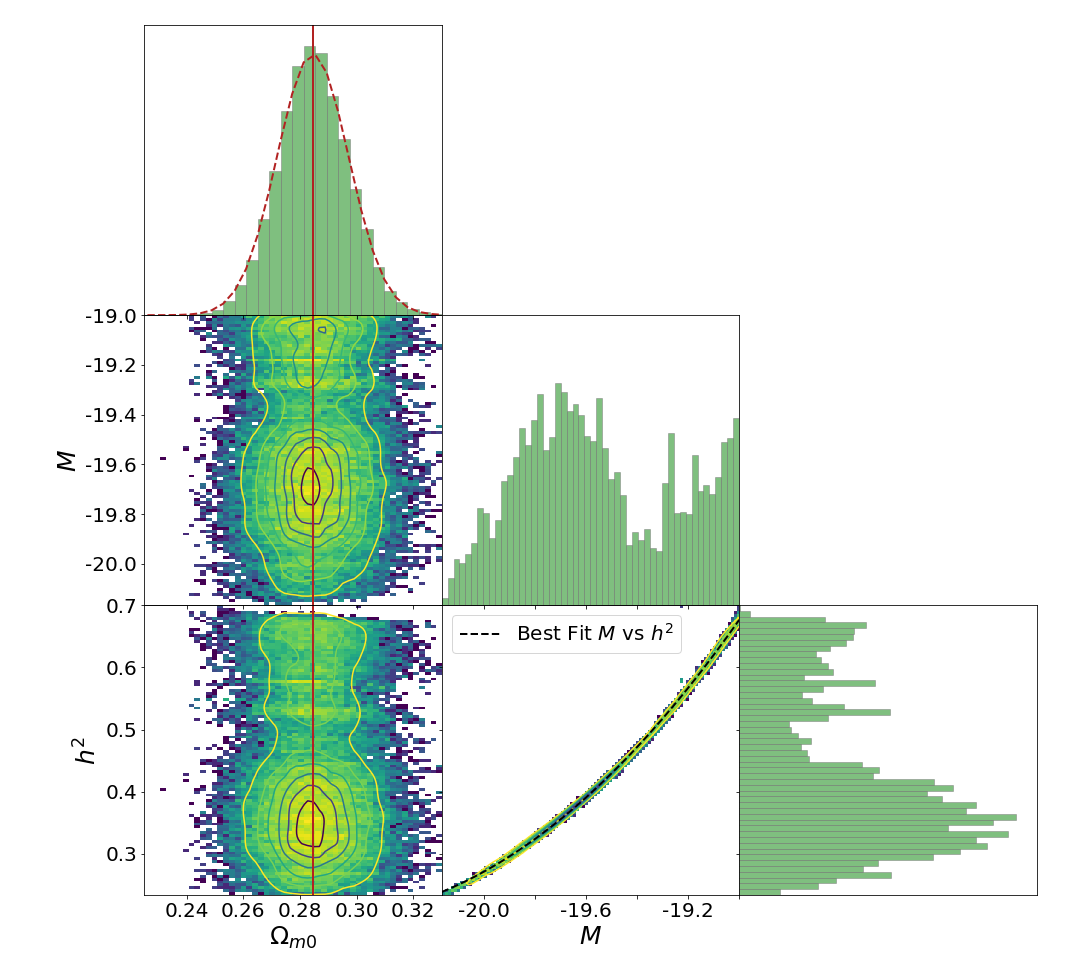}
\caption{}
\end{subfigure}%
\begin{subfigure}{.5\textwidth}
\centering
\includegraphics[width=\linewidth]{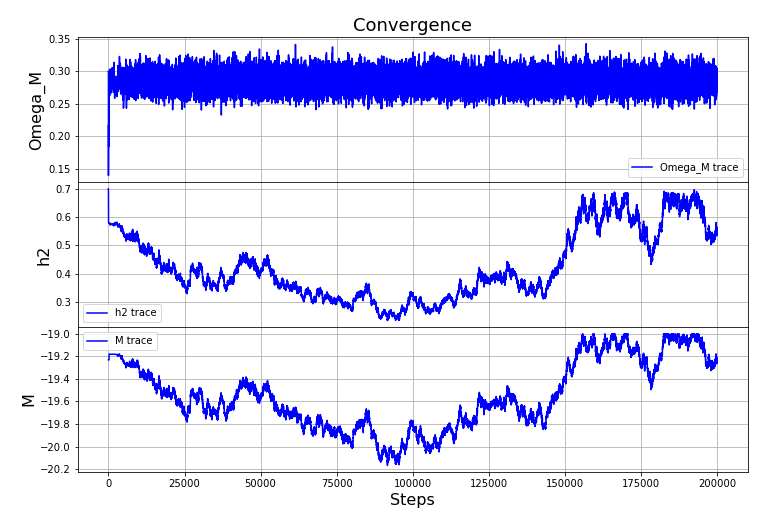}
\caption{}
\end{subfigure}
\caption{MCMC analysis assuming $M$ as a free parameter in GR. (\textbf{a})~Posterior probabilities densities. (\textbf{b})~Evolution of values with steps.}
\label{GRMfreeplots}
\end{figure}

For GR, we~did the same procedure, but~in this model the dependence appears between $h^2$ and $M$. The~polynomial is showed in Figure \ref{GRMfreeplots} and is given by:

\begin{equation}
h^2 = 0.177M^2 + 7.335 M + 75.896
\label{polinomioGR_M_C}
\end{equation}

Again, if~we evaluate Equation~\eqref{polinomioGR_M_C} at $M=-19.23 \pm 0.05$, we~obtain the $h^2$ value of previous~sections.

\unskip

\section{Conclusions}

Here we have studied the cosmological implications for a modified gravity theory, named Delta Gravity. The~results from SN-Ia analysis indicate that Delta Gravity explains the accelerating expansion of the Universe without $\Lambda$ or anything like ``Dark Energy''. The~acceleration is naturally produced by the Delta Gravity equations.

We assumed that $M_V=-19.23$ is a suitable value calculated from~\cite{Riess2016}. We~want to emphasize the very important fact that this value was obtained by local measurements and calibrations of SN-Ia, and~then, it~is independent from any cosmological model. Assuming~this, the~procedure presented does not use $\Lambda$CDM assumptions. We~only assume that the calibrations from Cepheids and SN-Ia are correct; therefore, the~absolute magnitude $M_V=-19.23$ for SN-Ia is reasonably correct. In~this case, the~Universe is accelerating, and this result is stable under any change of the priors for the MCMC analysis. Note~that the acceleration is highly determined by the $L_2$ value.

The acceleration in Delta Gravity is given by $L_2 \neq 0$. $L_2$ also determines that the Universe is made of Delta Matter and Delta Radiation. This~can be associated with the new field: $\tilde{\phi}$. It~is not clear if this Delta composition are real particles, or~not.

Also, Delta Gravity can predict a high value for $H_0$ (assuming $M_V = -19.23$). This~aspect is very important because the current $H_0$ value is in tension~\cite{Riess2016,Riess2018} between SN-Ia analysis and CMB Data. GR~also predicts a high $H_0$ value with the same assumptions, but~it needs to include $\Lambda$ to fit the SN-Ia Data. The~most important point about this, is~that the local measurement of $H_0$ is independent of the model.
\footnote{``The direct measurement is very model independent, but~prone to systematics related to local flows and the standard candle assumption. On~the other hand, the~indirect method is very robust and precise, but~relies completely on the underlying model to be correct. Any~disagreement between the two types of measurements could in principle point to a problem with the underlying \texorpdfstring{$\Lambda$}{L}CDM model.''~\cite{Odderskov}.} Furthermore, the~discrepancy about $H_0$ value could be indicating new physics beyond the Standard Cosmology Model Assumptions, and~maybe, one~possibility could be the modification of~GR.

Another difference between Delta Gravity and GR models, is~that Delta Gravity model predicts a Big Rip (as in phantom models~\cite{Robert2,Robert1}) that is dominated by the $L_2$ value. This~is shown in Figure \ref{DGEffectiveScaleFactorWithTime}.

The most important difference between Delta Gravity and the Standard cosmological model is the explanation about ``Dark Energy'' (the relation of $L_2$ with the accelerated expansion of the Universe).

\section*{Acknowledgements.}

The work of M. San Martin has been partially financed by Beca Doctorado Nacional Conicyt; Fondecyt 1150390;
CONICYT-PIA-ACT1417.J. Sureda has been partially financed by CONICYT-PIA-ACT1417; Fondecyt 1150390.The work of J. Alfaro is partially supported by Fondecyt 1150390, CONICYT-PIA-ACT1417.\\

\bibliographystyle{unsrt}
\bibliography{References}

\end{document}